\documentclass[aps,prl,amsmath,amssymb,amsfonts,floatfix,nofootinbib,superscriptaddress,longbibliography,12pt,reprint,noeprint]{revtex4-2}

\usepackage{geometry}
\usepackage[english]{babel}
\usepackage{blindtext}
\usepackage{amsmath,amsfonts,amssymb}

\usepackage{graphicx}
\usepackage{float}
\usepackage{times}
\usepackage{color}

\usepackage{subfigure}
\usepackage{indentfirst}
\usepackage{setspace}
\usepackage{bm}

\usepackage[colorlinks,linkcolor=blue,anchorcolor=blue,citecolor=green]{hyperref}
\usepackage{CJKutf8}
\newcommand{\Tr}{\mathrm{Tr}}

\graphicspath{{fig}}

\allowdisplaybreaks
\newcounter{amgg}
\newcounter{fan}
\newcounter{yc}

\begin{document}
\title{Measurement-Induced Integer-to-Half-Integer Entanglement Criticality in Non-Interacting Fermions with Random Long-Range Hopping}
\author{Can Yin (\begin{CJK*}{UTF8}{gbsn}殷灿\end{CJK*})}
\email{yin\_can@sjtu.edu.cn}
\affiliation{Shanghai Center for Complex Physics, School of Physics and Astronomy,
\\Shanghai Jiao Tong University, Shanghai 200240, China}
\affiliation{Tsung-Dao Lee Institute, Shanghai Jiao Tong University, Shanghai 201210, China}
\vspace{0.2cm}
\vspace{0.3cm}

\begin{abstract}
We study entanglement criticality of one-dimensional non-interacting fermions with random long-range hopping under continuous local-density monitoring via homodyne detection. To separate the heating and localizing effects of monitoring, we use the long-time stationary regime following an unmonitored unitary quench from a high-energy Slater-determinant state as an infinite-temperature reference. The entanglement transition points in the power-law hopping exponent $\alpha$ are $\alpha_c=1$ in the ground state, $2$ after the unmonitored quench, and $3/2$ in the long-time trajectory-averaged steady state under monitoring, giving the sequence $1\to2\to3/2$. The increase from $1$ to $2$ reflects the shift from Fermi-surface contributions to contributions from the full spectral distribution, 
while the subsequent decrease from $2$ to a value of $3/2$, which remains unchanged within the weak-monitoring regime, arises from the additional Euclidean time dimension in the effective superdiffusive nonlinear sigma model derived in the large bandwidth regime.
Large-scale GPU simulations of monitored quantum trajectories at large bandwidth $b=5$, where large system sizes are required to access the asymptotic scaling regime, enable us to reach system sizes up to $L\sim16384$, with finite-size scaling yielding $\alpha_c\approx1.6$ as well as a large effective correlation-length exponent $\nu$, supporting the analytical prediction.
\end{abstract}

\maketitle
\thispagestyle{article}

Non-interacting fermions with random long-range hopping can be described by the power-law random banded matrix (PRBM) model~\cite{MirlinEtAl1996}, and the spatial decay of the hopping strength controls quantum correlations and thus entanglement.
The same long-range disorder is known to drive an Anderson transition ~\cite{Anderson1958,EversMirlin2008} of single-particle states from extended metallic to localized insulating behavior~\cite{FyodorovMirlin1991,MirlinEtAl1996}, with the critical point set by the power-law decay exponent and characterized by universal finite-size scaling and multifractal eigenstates~\cite{AbrahamsEtAl1979,MirlinEvers2000,GarciaGarciaWang2005}, while the corresponding mobility edge marks the accompanying change in level correlations from quantum chaos to integrability~\cite{Mott1967,ShklovskiiEtAl1993}.
Such spectral signatures persist in many-body localization~\cite{AbaninEtAl2019}, as seen even in the many-body Fock space of an interacting quantum dot with no spatial dimension~\cite{AltshulerEtAl1997}, while free systems already exhibit nontrivial quantum correlations through the mixing of independently occupied orbitals~\cite{Peschel2003}.
In both single-particle and many-body settings, the effective nonlinear sigma model (NLSM)~\cite{Efetov1983,KamenevLevchenko2009} governing the infrared, with diffusive or superdiffusive soft modes, is generally expected to quantitatively capture the metal-insulator phase diagram.

With advances in quantum computing, monitored quantum many-body dynamics~\cite{Daley2014} has attracted growing interest, as measurement protocols extract information about a system~\cite{Bergquist1986,Murch2013} while the resulting backaction can be harnessed for qubit operations, precise control of quantum devices, and quantum-state preparation~\cite{WisemanMilburn1993,Minev2019,Sayrin2011}.
Continuous measurement along real-time quantum trajectories can either enhance or reduce entanglement, while its average effect is to suppress quantum correlations~\cite{LeGal2024,Fan2026SingleMeasurement}, as illustrated in a simple setting of 1D free spinless fermions with nearest-neighbor hopping, where arbitrarily weak measurement suppresses the spreading of entanglement and drives the fully scrambled long-time state into an area-law localized phase~\cite{Poboiko2023,Fan2026GPU}.

The power-law exponent $\alpha$ for the PRBM $H_{ij}$, with variance $\overline{|H_{ij}|^2}\sim r^{-2\alpha}$~\cite{FyodorovMirlin1991}, sets the regimes of extended and localized phases and is therefore a useful probe for identifying a measurement-induced phase transition (MIPT)~\cite{LiChenFisher2018,SkinnerRuhmanNahum2019,Noel2022,Hoke2023,Koh2023} at $\alpha_c$.
For slowly decaying hopping, continuous monitoring cannot fully suppress the long-range quantum correlations, as the system remains RMT-like for $\alpha<\frac{1}{2}$ due to the divergence of $\sum_r r^{-2\alpha}$ and typically exhibits extended metallic behavior for $\frac{1}{2}<\alpha<1$~\cite{MirlinEtAl1996}.
At larger $\alpha$, it has been shown that, under weak measurement, monitored long-range hopping models with deterministic power-law amplitudes exhibit no localized entanglement phase for $1<\alpha<\frac{3}{2}$~\cite{Minato2022,Muller2022}, and previous work on the same monitored PRBM model investigated in this paper suggests no extended sub-volume-law phase for $\alpha\gtrsim\frac{3}{2}$ and, based on numerical finite-size crossings at bandwidth $b=1$, a MIPT at $\alpha_c\approx\frac{3}{2}$~\cite{Fan2026LongRange}, whereas increasing the measurement rate toward the hopping scale shifts the transition point downward in both models to $\alpha_c\in(1,\frac{3}{2})$, as expected from the direct competition between monitoring and long-range hopping.

It is known that continuous monitoring pumps energy into a closed system and thus drives its real-time nonequilibrium dynamics toward a long-time steady state, which corresponds to infinite temperature in the maximally mixed trajectory-averaged ensemble within the relevant conserved-charge sector~\cite{Gurvitz2000,YiEtAl2011,BuchholdEtAl2021,Poboiko2023,GuoEtAl2025}. Previous works have established universality in the monitored dynamics of short-range fermions, with the localization pattern determined by the Hamiltonian symmetry through the corresponding NLSM of class AIII for on-site disorder and BDI for the clean reference case~\cite{Poboiko2025Interacting,YinEtAl2026Disorder}, and this AIII universality naturally extends to the PRBM in the fast-decay regime $\alpha>\frac{3}{2}$, where the same diffusive short-distance dynamics is typically recovered, implying that, in the weak-measurement regime, the entanglement-localization pattern remains fully controlled by disorder and that the MIPT occurs at a constant critical exponent $\alpha_c$.

The shift of $\alpha_c$ relative to the Anderson transition at $\alpha=1$~\cite{MirlinEtAl1996} originates from a measurement-induced dimensional effect. An early route was based on representing the spacetime history of real-time evolution in $d$-dimensional systems with local Hamiltonian dynamics as a tensor network in $d+1$ dimensions, and then mapping this network to a random free-fermion system in $d+1$ dimensions by interpreting the stochasticity introduced by monitoring along the temporal direction as disorder, thereby capturing certain features of monitored dynamics~\cite{SkinnerEtAl2019,VasseurEtAl2019,JianEtAl2020,LiEtAl2021,JianEtAl2022}. It later turned out that such a higher-dimensional background picture is heuristic rather than a controlled microscopic description, and that the dimensional effect is manifested only in the perturbative soft-mode sector of the effective theory, described by a $d+1$-dimensional NLSM whose target manifold distinguishes universality classes~\cite{FavaEtAl2023,NahumWiese2023,FavaEtAl2024,Poboiko2023,PoboikoEtAl2024}, with a systematic microscopic route based on a replicated Keldysh path integral, the replica limit $R\to1$ implementing the Born weighting, and saddle-point analysis extracting entanglement entropy (EE) from the leading soft fluctuations~\cite{Poboiko2023,GuoEtAl2025}. 

In this work, following this microscopic route, we find that the monitored PRBM is described by a higher-dimensional extension of the superdiffusive NLSM for Anderson localization, which shifts the critical power-law exponent $\alpha_c$ by $\frac{1}{2}$ to $\frac{3}{2}$, with this shift arising from the competition between the enhancement from the ground state to the infinite-temperature steady state and the localizing effect of monitoring. To show this, we first consider the non-monitoring case of the quench dynamics of an initial Néel state under long-time unitary evolution, and find that the transition point shifts from the analytically derived ground-state value $\alpha_c=1$ to $\alpha_c=2$, where long-range hopping is suppressed and becomes subleading, while the critical scaling is interpreted in terms of rare two-level resonances~\cite{Levitov1990} arising from the weak long-distance hopping. Then we add continuous homodyne measurement to the same Néel initial state, follow the long-time non-unitary evolution of quantum trajectory ensembles, analytically construct the replicated Keldysh path integral and map entanglement dynamics onto the superdiffusive NLSM with an extra Euclidean temporal dimension, with the tree-level RG flow of this NLSM yielding $\alpha_c=\frac{3}{2}$.

{\it The model.-} We consider a one-dimensional non-interacting fermionic chain with power-law random hopping, whose hopping matrix is a real symmetric PRBM~\cite{MirlinEvers2000},
\begin{equation}
H=\sum_{i,j=1}^{L}h_{ij}\hat{c}_i^\dagger \hat{c}_j, \qquad h_{ij}=\frac{J\xi_{ij}}{\sqrt{1+(d_{ij}/b)^{2\alpha}}},
\label{eq:PRBM_Ham}
\end{equation}
where $\hat{c}_i$ and $\hat{c}_i^\dagger$ are the annihilation and creation operators for fermions at sites $i=1,2,\cdots,L$, $d_{ij}=\frac{L}{\pi}\left|\sin\frac{\pi(i-j)}{L}\right|$ is the distance between sites $i$ and $j$ under periodic boundary conditions, and $\xi_{ij}$ are independent Gaussian random variables for $i\leq j$, satisfying $\xi_{ij}=\xi_{ji}$ with zero mean and unit variance. Here, $J$ sets the overall hopping magnitude and is fixed to $J=1$, $b$ sets the bandwidth and $\alpha$ controls the power-law decay of the hopping variance. We choose the half-filled Néel state $|\psi\rangle_0=|010101\cdots\rangle$ as the initial state, with $0$ and $1$ denoting the site occupation numbers. We take the large-bandwidth limit $b\to\infty$ in the analytical path-integral formalism and $b=5$ in numerical simulations. Each realization of the Hamiltonian remains fixed during the time evolution.

For the monitored dynamics, we employ homodyne measurement of the local particle density $n_i=c_i^\dagger c_i$ at each site $i=1,2,\cdots,L$, with measurement rate $\gamma$. The state evolves along each quantum trajectory according to the normalized stochastic Schrödinger equation driven by Gaussian noise, and the particle number $N=\frac{L}{2}$ is conserved along each quantum trajectory,
\begin{equation}
\begin{aligned}
d|\psi(t)\rangle={}&\Bigl[-iH\,dt-\frac{\gamma}{2}\sum_{i=1}^{L}\bigl(n_i-\langle n_i\rangle_t\bigr)^2dt\\
&+\sum_{i=1}^{L}\bigl(n_i-\langle n_i\rangle_t\bigr)dW_i^t\Bigr]|\psi(t)\rangle.
\end{aligned}
\label{eq:PRBM_QSD}
\end{equation}
where $\langle n_i\rangle_t=\langle\psi(t)|n_i|\psi(t)\rangle$, and the real Wiener increments $dW_i^t$ are independent across sites and times, with zero mean and variance $\gamma dt$. At each finite time step $\Delta t$, all sites are weakly measured, introducing an additional normalization factor for each quantum trajectory, and we perform a QR decomposition numerically for normalization at each step, as detailed in the End Matter. We consider the long-time trajectory-averaged steady state, which is maximally mixed, and numerically evolve the state until its EE reaches the fully scrambled regime before averaging over equally spaced time points.

{\it Numerical Methods and Performance.-} We implement the numerical evolution using C++ and CUDA, initializing all data and performing nearly all computations on the GPU, employing custom CUDA kernels for selected performance-critical operations to reduce kernel-launch overhead and avoid unnecessary synchronization, and using cuBLAS, cuSOLVER, and cuSPARSE for standard linear-algebra operations. For the monitoring case, we evolve the state using a fourth-order Runge-Kutta method with a time step of $dt=0.05$ and computational complexity $O(L^4)$, reaching $L=16384$ (168 trajectories in total) at bandwidth $b=5$, about five times the largest size $L=3200$ studied at $b=1$~\cite{Fan2026LongRange}, a system size necessary to control finite-size effects that scale roughly with $b$ and remain significant for $L< 4096$. For monitoring rates $\gamma=0.3,0.5,0.7$ satisfying $\gamma\ll J\sqrt{b}$ and corresponding to the weak-monitoring end of the range $\gamma\in[0.1,8]$ studied at $b=1$ in~\cite{Fan2026LongRange} after accounting for the $1/\sqrt{b}$ scaling of the effective monitoring rate, we find that the transition point remains near $\alpha_c\approx1.6$, complementing the broader measurement-rate dependence reported in that work. The entire computation took approximately 3.5 months using three A100 and seven V100 GPUs.
For the non-monitoring case, we avoid cumulative time evolution by directly computing the state at each selected time, with the computational cost dominated by eigenvalue problems scaling as $O(L^3)$, allowing us to reach $L=24336$ (4400 disorder realizations in total, with 10 time points each) in about one week on the same GPU workstations used for the monitoring case.

{\it Non-monitoring entanglement transition.-} For the single-particle PRBM with large bandwidth $b\gg 1$, wavefunction localization is governed by the (super)diffusive pole of the response kernel $\Gamma(q,\omega)\sim (Dq^{2\alpha-1}-i\omega)^{-1}$, where $D$ is the diffusion coefficient and $\omega$ is the response frequency, and an Anderson transition occurs at $\alpha_c=1$ with multifractality~\cite{MirlinEtAl1996,FyodorovMirlin1991}. The non-interacting many-body PRBM has the same single-particle representation, with a filling fraction $n=\frac{N}{L}$ and a Fock-space many-body wavefunction. The ground-state EE $S_A$ in Eq.~\ref{eq:EE} can be approximated by the leading-order particle-number cumulant $\langle\langle N^2\rangle\rangle$ as $S_A\approx \frac{\pi^2}{3}\langle\langle N^2\rangle\rangle$~\cite{KlichLevitov2009}, which is determined by particle-hole excitations across the Fermi energy $E_F$~\cite{Burmistrov2017}, $E\in[E_F-\frac{\omega}{2},E_F+\frac{\omega}{2}]$, and governed by the same superdiffusive kernel $\Gamma(q,\omega)$. 
The similarity between ground-state many-body entanglement and the single-particle Anderson transition is expected from their common PRBM ensemble, and we analytically derive the entanglement transition at $\alpha_c=1$ by relating the EE to $\Gamma(q,\omega)$, and identify the critical scaling $S_A\sim(\log L)^2$, as detailed in Section 1 of the SM. A previous numerical study also reported the transition at $\alpha=1$, but the fitted power law $S_A\sim L^{0.33}$ does not correctly characterize the critical scaling~\cite{PouranvariYang2014}. More recently, it was numerically found that the transition near $\alpha=1$ exhibits the expected critical marginal behavior, with $S_A\sim\log L$~\cite{FanGarciaGarcia2026}, while the distinction between the numerically observed logarithmic scaling and our predicted $(\log L)^2$ scaling is difficult to resolve numerically at finite system sizes.

\begin{figure}[!htb]
	\centering
	\includegraphics[width=7.5cm]{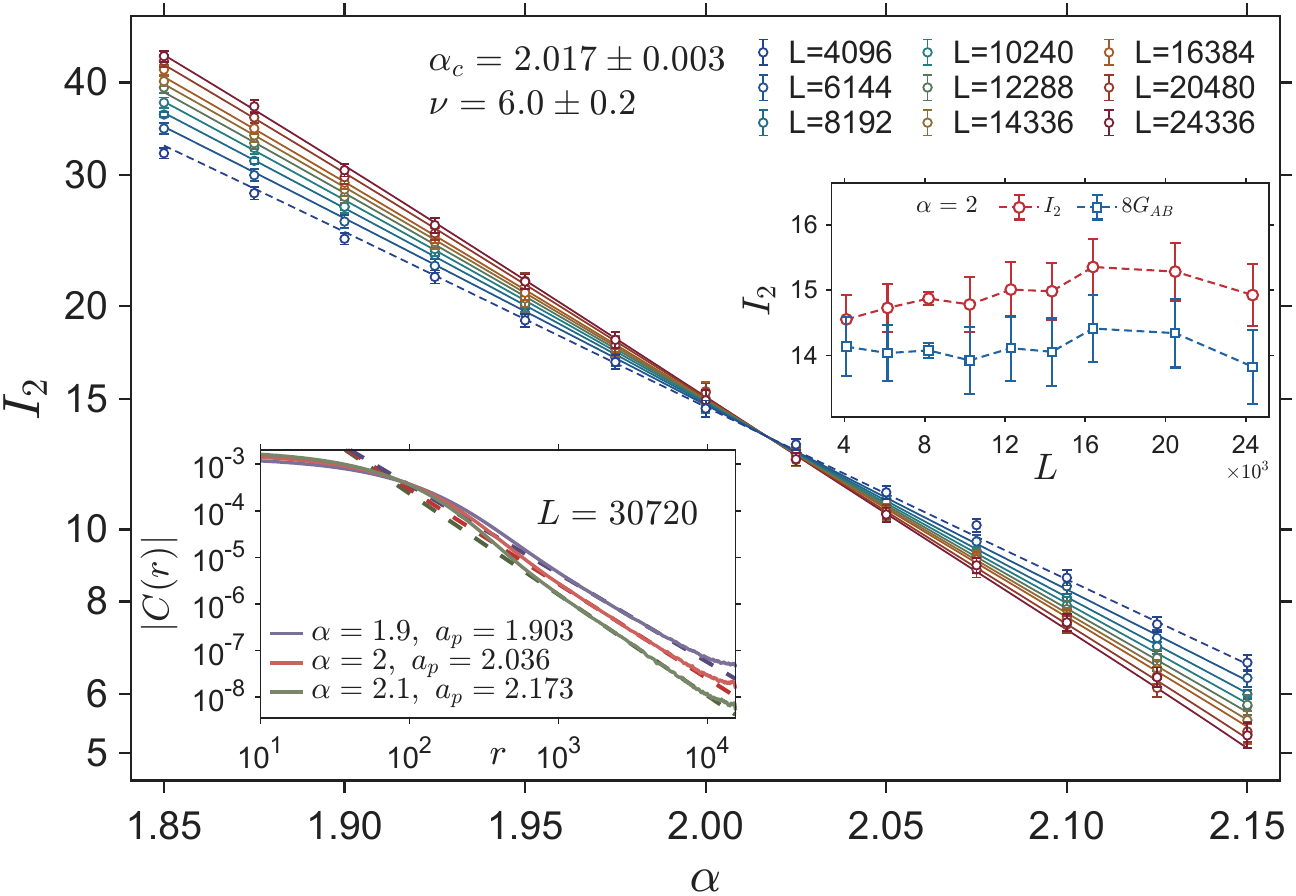}
	\caption{Mutual information $I_2$ in the steady state of unitary dynamics starting from an initial Néel state, for system sizes $L\in[4096,24336]$ and hopping exponents $\alpha\in[1.85,2.15]$. Error bars indicate two-sided $95\%$ Student's $t$ confidence intervals for the mean, computed by pooling all individual trajectories with equal weight across disorder realizations, without intermediate averaging. At least 100 disorder realizations are used for each $(L,\alpha)$, with observables evaluated at ten equally spaced late times in $[10^3L,10^4L]$ for each realization, after the observables have reached their long-time stationary regime.
The solid curves show a joint finite-size scaling fit to $I_2$ data for $L\geq6144$, using a quadratic relevant scaling field ($m=2$) and a cubic scaling function ($n=3$). The fit yields a common crossing at $\alpha_c=2.017\pm0.003$ and a correlation-length exponent $\nu=6.0\pm0.2$, where the uncertainties indicated by $\pm$ represent one standard deviation.
 The dashed curve shows the prediction from the same fit for $L=4096$, with the corresponding data excluded from the fit.
\textbf{(Lower-left inset)} Log--log plot of the absolute value $|C(r)|$ at $L=30720$ for different $\alpha$. Error bars are omitted as they are generally narrower than the plotted lines at the displayed scale. The solid curves show geometric averages taken first over sampled times and then over trajectories across disorder realizations. The dashed lines show power-law fits over selected tail intervals. \textbf{(Upper-right inset)} Comparison of $I_2$ and $8G_{AB}$ at $\alpha=2$ for all system sizes. The difference remains small across the sizes shown, and both properties show no clear system-size dependence.
}\label{fig:nonmoncross}
\end{figure}

Monitoring continuously heats a particle-number-conserving system toward infinite temperature while inducing localization, so, to separate these effects, we consider the long-time unitary evolution of a high-energy initial Slater-determinant state, whose disorder-averaged steady state reproduces the same maximally mixed infinite-temperature state as the monitored trajectory average.
For the half-filled initial N\'eel state $|\psi\rangle_0$, we plot in Fig.~\ref{fig:nonmoncross} the mutual information $I_2$ defined in Eq.~\ref{eq:I2} as a function of the hopping exponent $\alpha\in[1.85,2.15]$ for different system sizes $L\in[4096,24336]$, and use its scale independence at criticality to locate the entanglement transition through finite-size crossings.
To determine the critical point, we perform a finite-size scaling analysis by defining $\phi=u(w)L^{1/\nu}$, where $w=(\alpha-\alpha_c)/\alpha_c$ and $u(w)=\sum_{j=1}^{m}b_jw^j$, and expanding $I_2=\sum_{i=0}^{n}a_i\phi^i$, using polynomial orders $m=2$ and $n=3$, fixing $a_1=1$ and performing weighted least-squares regression with inverse-squared SEM weights~\cite{SlevinOhtsuki2014}, yielding a clear entanglement transition at $\alpha_c=2.017\pm0.003\approx2$, in contrast to the ground-state transition at $\alpha_c=1$.
The density-density correlation function $C(r)=\overline{\langle\delta n_i\delta n_j\rangle}$, with $r=|i-j|$, exhibits $C(r)\sim r^{-\alpha}$, as shown in the lower-left inset of Fig.~\ref{fig:nonmoncross}, implying $I_2\sim L^{2-\alpha}$ and a logarithmic scaling variable $(\alpha-\alpha_c)\log L$ near $\alpha_c=2$, so that fitting with a scaling function of $(\alpha-\alpha_c)L^{1/\nu}$ yields an effective $\nu$ that increases with $L$ and diverges as $L\to\infty$, consistent with the large fitted value $\nu=6.0\pm0.2$.
The result is surprising because, although heating is expected to increase dynamical freedom and may lift the localized phase relative to the ground state, the single-particle metal-insulator transition is fixed by the PRBM structure and independent of the many-body state, and for $\alpha\geq\frac32$ the leading diffusion propagator $\Gamma(q,\omega)\sim(Dq^2-i\omega)^{-1}$ is independent of $\alpha$.

This motivates us to examine rare long-range resonances as a possible mechanism for the entanglement transition. Assuming that the transition occurs in the localized PRBM regime $\alpha>\frac32$ at large bandwidth $b\gg1$, we identify the parameter $\eta$ as the expected number of direct resonant partners of a localized orbital beyond its localization length, which scales as $b^{\frac32-\alpha}$ and vanishes, allowing the entanglement to be described by rare independent resonance pairs. The direct calculation in Sec.~2 of the Supplemental Material, based on two-level resonances, yields $C(r)\sim r^{-\alpha}$ for $\alpha>\frac32$, supporting an entanglement transition at $\alpha_c=2$, and also gives $\frac{I_2}{G_{AB}}=8$ instead of the $\frac{2\pi^2}{3}$ coefficient of the leading-order cumulant expansion, as supported by the upper-right inset of Fig.~\ref{fig:nonmoncross}, where $I_2$ and $8G_{AB}$ differ only slightly, while the entanglement transition point remains determined by the divergence of the distance-weighted correlation sum $\sum_r rC(r)$. Since long-range hopping becomes stronger for $\alpha\leq\frac32$, we expect only a single transition at $\alpha_c=2$. The same resonance analysis also applies to the $b\ll1$ regime, suggesting that $\alpha_c=2$ is the entanglement transition point across the full range of bandwidths $b$.

{\it Measurement-induced entanglement transition.-} We consider entanglement dynamics under continuous homodyne monitoring, following the quantum trajectory evolution in Eq.~\ref{eq:PRBM_QSD}. Analytically, we construct an $R$-replica Keldysh path integral based on the cumulant expansion of EE~\cite{KlichLevitov2009}, using the replica limit $R\to1$~\cite{Poboiko2023} to cancel the Born weight associated with each measurement event, as detailed in Sec.~3 of the Supplemental Material. We first derive the Grassmann action $iS[\bar{\Psi},\Psi]$ from the master equation for the replicated density matrix $\rho^R\equiv\overline{\rho^{\otimes R}}$, as Gaussian factorization into two-point contractions of the non-Markovian temporal correlations induced by static power-law hopping disorder allows a Lindbladian-form representation. We then express the action in the doubled basis $\Phi^\dagger=(\bar\Psi,-\Psi)$, $\Phi=(\Psi,\bar\Psi)$ as $iS[\Phi^\dag,\Phi]$ to combine the direct and exchanged disorder contractions $\overline{h_{ij}h_{ij}}$ and $\overline{h_{ij}h_{ji}}$ for lattice sites $i,j$,
\begin{equation}
\begin{aligned}
iS[&\Phi^\dagger,\Phi]=\Tr\Bigg[\frac{i}{2}\sum_i\Phi_i^\dagger
\left(\begin{smallmatrix}
i\sigma_K^z\partial_t-0^+\sigma_K^y&0\\
0&-i\sigma_K^z\partial_t+0^+\sigma_K^y
\end{smallmatrix}\right)\Phi_i\\
-&\frac{\gamma}{8}\delta(t-t')\sum_i(\Phi_i^\dagger\Phi_i)^2+\frac{1}{8}\sum_{ij}a(|i-j|)(\Phi_i^\dagger\sigma^z\Phi_i)\\
(&\Phi_j^\dagger\sigma^z\Phi_j)+\frac{\gamma(R-1)}{2}\delta(t-t')\sum_i(1-2\langle n_i\rangle_t)(\Phi_i^\dagger\Phi_i)\Bigg]
\end{aligned}
\label{eq:Phiaction}
\end{equation}
where we have used $n=1/2$, so that the Keldysh distribution $F=1-2n$ encoded in the regulator of the bare quadratic kernel vanishes in the infinite-temperature steady state. $\Tr$ denotes the matrix trace over the doubled $4R$-dimensional replica-Keldysh space and the convolution over the two time indices $t,t'$, $\sigma_K^{z,y}$ are the Pauli matrices in $2\times 2$ Keldysh space, $0^+$ is an infinitesimal regulator, $\delta(t-t')$ restricts the monitoring terms to equal times, while the static disorder term couples two independent time points, and the nonlocal spatial kernel $a(|i-j|)=\frac{1}{1+(d_{ij}/b)^{2\alpha}}$ represents the hopping variance in Eq.~\ref{eq:PRBM_Ham}.

The entanglement dynamics is governed by leading-order fluctuations around the saddle point of the action~\cite{Poboiko2023} in Eq.~\ref{eq:Phiaction}. To determine this saddle point, we introduce the collective field $\mathcal G=i\Phi\Phi^\dagger$ and the associated field $Q$ through a generalized Hubbard--Stratonovich transformation to represent the action as $iS[\mathcal G,Q]$, retaining the two leading low-momentum pairing channels of the quartic interactions, corresponding to the matrix-trace $\Tr(\mathcal G^2)$ and trace-square $(\Tr\mathcal G)^2$ quadratic structures, and integrating out $\mathcal G$ to obtain the action $iS[Q]$. We assume a replica-symmetric saddle at $R=1$ that is stationary, time-local ($t=t'$), and spatially homogeneous, and then perform the Keldysh rotation to impose the causality condition $\Tr(Q)=0$ in the rotated basis.
At half filling, the vanishing Keldysh distribution yields $\Tr(Q\sigma_K^x)=0$.
For the localized but short-time superdiffusive regime of the PRBM ensemble with $\alpha\in (1,\frac{3}{2})$~\cite{MirlinEtAl1996}, the power-law tail of the hopping variance yields, after the momentum-independent contribution is subtracted, the nonanalytic spatial kernel $a_{k=0}-a_k=2\sum_{r=1}^{\infty}a(r)(1-\cos kr)\approx C_\alpha b^{2\alpha}|k|^{2\alpha-1}$, which enters the action $iS[Q]$ below:
\begin{equation}\label{eq:SQ}
\begin{aligned}
iS[Q]\approx&\frac12\operatorname{Tr}\Bigg[\sum_i\ln\left(K_0+\frac{i(\gamma+a_{k=0})}{4}Q\right)_{ii}\\
&-\frac{C_\alpha b^{2\alpha}}{16}\sum_k|k|^{2\alpha-1}Q_kQ_{-k}\Bigg],
\end{aligned}
\end{equation}
where $C_\alpha=2\int_0^\infty dx \frac{1-\cos x}{x^{2\alpha}}$, and $K_0$ denotes the first free quadratic term appearing in Eq.~\ref{eq:Phiaction}. We also neglect the massive time-bilocal terms with $t\neq t'$, which do not contribute at leading order in the infrared. Varying Eq.~\ref{eq:SQ} yields the standard saddle $\Lambda=\sigma_K^z$, and the leading-order fluctuations in the replicon sector for general replica number $R$, parametrized by a slowly varying field $U$, give rise to the superdiffusive NLSM action $iS[U]$, 
\begin{equation}\label{eq:SU}
iS[U]\approx-g\operatorname{Tr}\sum_k\left(|k|^{2\alpha-1}+\tilde\Omega^2\right)U_{k,\tilde\Omega}(U^\dagger)_{-k,-\tilde\Omega},
\end{equation}
where $U$ takes values in the target space $SU(R)$, as follows from the coset structure of the Goldstone modes, $[SU(R)\times SU(R)]/SU(R)\simeq SU(R)$\cite{Poboiko2023,Poboiko2025Interacting,YinEtAl2026Disorder}. The rescaled time $\tilde t$ and the bare stiffness constant $g$ are defined as $\tilde t=\frac{b^\alpha}{2}\sqrt{C_\alpha(\gamma+a_{k=0})}t$ and $g=\frac{b^\alpha}{4}\sqrt{\frac{C_\alpha}{\gamma+a_{k=0}}}$, respectively. Here, $\tilde\Omega$ denotes the external transfer frequency conjugate to the rescaled time $\tilde t$ of the slowly varying field $U$, and the trace $\Tr$ also includes integration over $\tilde\Omega$.

Under the anisotropic rescaling $x\to sx$ and $\tilde t\to s^z\tilde t$, with the dynamical exponent $z=\alpha-\frac12$, the stiffness scales at tree level as $g\to s^{1-z}g$, yielding
\begin{equation}
\frac{dg}{d\log l}=\left(\frac32-\alpha\right)g,
\end{equation}
where $l$ denotes the running RG length. For $1<\alpha<\frac32$, the positive RG coefficient drives the stiffness toward larger values, corresponding to a flow toward the weak-coupling regime of the NLSM. Correspondingly, both the mutual information $I_2$ and the leading cumulant $G_{AB}$ increase with system size. At $\alpha=\frac32$, the stiffness becomes marginal at tree level, identifying the upper boundary of the superdiffusive regime. This boundary is reached continuously from within $1<\alpha<\frac32$, as the spatial kernel approaches a quadratic momentum dependence. At this boundary, $C(r)\sim r^{-2}$ renders $G_{AB}$, and hence $I_2$, scale invariant, as the two spatial integrations compensate for the $L^{-2}$ decay of $C(r)$. For $\alpha>\frac32$, the negative RG coefficient drives the stiffness toward smaller values, corresponding to a flow toward the strong-coupling regime of the NLSM, and, because the two regions become asymptotically uncorrelated, both $G_{AB}$ and $I_2$ decrease with increasing system size.

\begin{figure}[!htb]
	\centering
	\includegraphics[width=7.5cm]{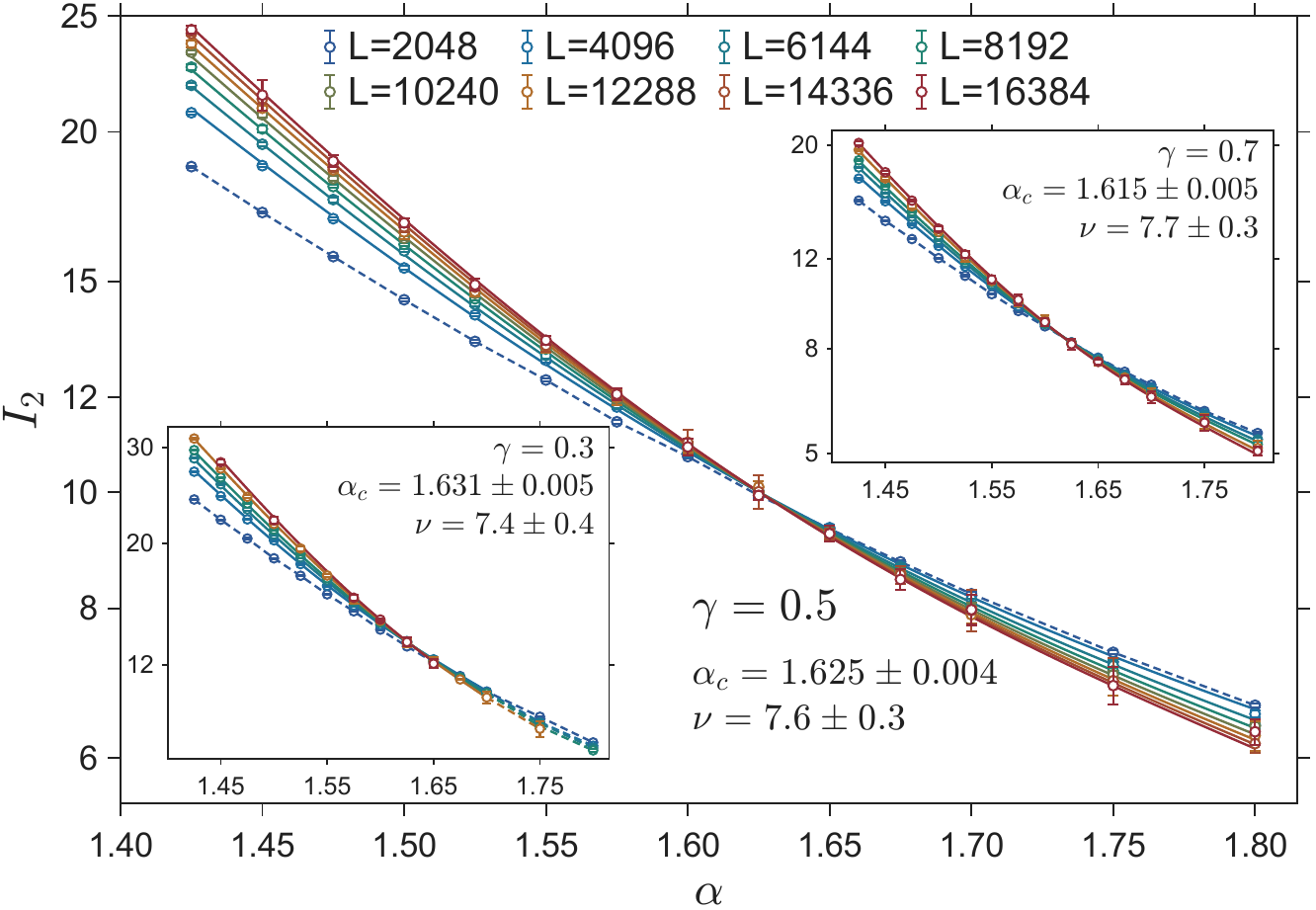}
	\caption{Mutual information $I_2$ in the long-time steady state under continuous monitoring, starting from an initial Néel state, at monitoring rates $\gamma=0.3,0.5,0.7$ (\textbf{Main plot}: $\gamma=0.5$; \textbf{Lower-left inset}: $\gamma=0.3$; \textbf{Upper-right inset}: $\gamma=0.7$), for system sizes $L\in[2048,16384]$ and hopping exponents $\alpha\in[1.425,1.8]$. Error bars indicate two-sided $95\%$ Student-$t$ confidence intervals, computed by pooling individual trajectories across all disorder realizations. The dashed curves connect the data points for $L=2048$, with the corresponding data excluded from the fit. The solid curves show joint polynomial finite-size scaling fits to $I_2$ data for $L\geq4096$, using $m=2$ and $n=3$. The fits yield crossings at $\alpha_c=1.631\pm0.005$ ($\gamma=0.3$), $\alpha_c=1.625\pm0.004$ ($\gamma=0.5$), and $\alpha_c=1.615\pm0.005$ ($\gamma=0.7$), with correlation-length exponents $\nu=7.4\pm0.4$ ($\gamma=0.3$), $\nu=7.6\pm0.3$ ($\gamma=0.5$), and $\nu=7.7\pm0.3$ ($\gamma=0.7$), where the uncertainties indicated by $\pm$ denote one standard deviation. The crossing points remain stable around $\alpha_c\approx1.6$, while the correlation-length exponent $\nu$ remains large and increases with system size, consistent with the analytical prediction.
}\label{fig:moncross}
\end{figure}
In Fig.~\ref{fig:moncross}, we plot the mutual information $I_2$ as a function of $\alpha$ for different system sizes $L$, averaging over trajectories across disorder realizations in the long-time scrambled regime as in Fig.~\ref{fig:monitorEEL}(left) of the End Matter. The crossing marks a reversal in size dependence, with $I_2$ increasing with $L$ in the subvolume-law phase and decreasing with $L$ in the area-law phase, as shown in Fig.~\ref{fig:monitorI2GAB}(left) of the End Matter. The large bandwidth $b=5$ enhances finite-size effects, so we perform finite-size scaling using GPU simulations for $L\in [4096, 16384]$. Even our smallest fitted size exceeds the largest size ($L=3200$) used in the $b=1$ crossing analysis~\cite{Fan2026LongRange}. For $\gamma=0.3,0.5,0.7$, we obtain $\alpha_c=1.631(5)$, $1.625(4)$, and $1.615(5)$, respectively. The crossing varies little with $\gamma$, unlike the strong $\gamma$ dependence reported in the Zeno regime at $b=1$~\cite{Fan2026LongRange}. This weak dependence supports the analytical prediction that $\alpha_c=\frac32$ is independent of $\gamma$. We attribute the offset of about $0.1$ to finite-size effects. As shown in Fig.~\ref{fig:monitorCrap} of the End Matter, for both $\gamma=0.5$ and $0.7$, the decay exponent $a_p$ in $C(r)\sim r^{-a_p}$ increases with $L$ at fixed $\alpha$, so that $a_p\approx2$ is reached at smaller $\alpha$ for larger $L$. At fixed $L$, $a_p(\alpha)$ continues to increase for $\alpha>\alpha_c$, but its slope decreases as it approaches $2$. This bending likely reflects the competition between localization and short-range propagation described by the NLSM kernel $g(|k|^{2\alpha-1}+\tilde\Omega^2)$ in Eq.~\ref{eq:SU}.

{\it Conclusions.-} We have investigated the entanglement transition from the subvolume-law phase to the area-law phase in the one-dimensional PRBM model through analytical and numerical studies of the unmonitored high-energy N\'eel-state quench and the continuously monitored quantum-trajectory dynamics. Without monitoring, our two-level resonance analysis of the long-time N\'eel-state quench yields $C(r)\sim r^{-\alpha}$ for $\alpha>\frac32$ and predicts the transition to shift from the analytically derived ground-state value $\alpha_c=1$ to $2$, a result supported by numerical simulations up to $L=30720$. Under continuous monitoring, the replicated Keldysh field theory leads to a superdiffusive NLSM whose RG flow predicts that $\alpha_c=\frac32$ is independent of the monitoring rate, consistent with large-scale GPU simulations of the monitored dynamics up to $L=16384$, which yield crossings near $\alpha_c\approx1.6$ at several weak monitoring rates. Together, our analytical theory and large-scale numerical simulations establish the mechanism underlying the measurement-induced shift in entanglement criticality.

\section{End Matter}
In this End Matter, we detail the numerical methods for simulating the entanglement dynamics of the PRBM under homodyne monitoring and provide additional numerical results to support the validity and robustness of main-text findings.

Numerically, the quantum trajectory evolution preserves Gaussianity even in the presence of monitoring, so Wick's theorem allows us to express any observable in terms of the $L\times L$ correlation matrix $D$, defined as $D_{ij}(t)\equiv\langle\psi(t)|c_j^\dagger c_i|\psi(t)\rangle$. We represent the Slater determinant state $|\psi(t)\rangle$ by an $L\times N$ matrix $U$ of occupied orbitals, such that 
\begin{equation}\label{eq:SlaterUmatrix}
|\psi(t)\rangle = \prod_{k=1}^{N} \Bigg[\sum_{j=1}^L U_{jk}(t) c_j^\dagger\Bigg] | \text{vac} \rangle,
\end{equation}
where $|\text{vac} \rangle$ denotes the vacuum state, $U^\dagger U=1$, and $D=UU^\dagger$. After each time step $t\rightarrow t+dt$, we perform a QR decomposition $U=QR$ and replace $U$ with $Q$, where $Q^\dagger Q=\mathbf1_N$ and $R$ is upper triangular. We evolve $U(t)$ using the fourth-order Runge–Kutta method, whose numerical error increases with $\gamma$ and $dt$ due to the leading $O(\sqrt{\gamma dt})$ scaling of $dW_i^t$, and we have verified that the accumulated numerical error at each data point is negligible compared with its statistical variance. We fix $dt=0.05$ across all monitoring rates because, although a smaller $dt$ reduces the error at larger monitoring rates, it also leads to more frequent and weaker measurements, while the resulting $dt$ dependence of entanglement remains distinguishable from both numerical integration errors and statistical fluctuations.

The entanglement entropy (EE) of the half-chain $A=\{1,2,\ldots,L/2\}$, with the cut at the center, is $S_A(t)=-\operatorname{Tr}_A[\rho_A(t)\log\rho_A(t)]$, where $\rho_A(t)=\operatorname{Tr}_{\bar A}\rho(t)$ is the reduced density matrix and $\rho(t)=|\psi(t)\rangle\langle\psi(t)|$.
For free fermions, 
\begin{equation}\label{eq:EE}
S_A(t)=-\sum_{a=1}^{\frac{L}{2}}\bigl[\lambda_a(t)\log\lambda_a(t)+(1-\lambda_a(t))\log(1-\lambda_a(t))\bigr],
\end{equation}
where $\lambda_a(t)$ are the eigenvalues of $D_A(t)$, the restriction of $D(t)$ to sites in $A$. 

Analytically, it is more convenient to expand the EE in a series of cumulants of the particle number. Perturbatively, it was found that only the second cumulant contributes at leading order, yielding 
\begin{equation}
S_A(t)\approx\frac{\pi^2}{3}\int_0^{L/2}dx\int_0^{L/2}dy C(x-y,t),
\end{equation}
where $C(x-y,t)$ is the connected density-density correlation function, whose lattice expression follows from Wick's theorem as $C_{ij}(t)=\langle n_i n_j\rangle_t-\langle n_i\rangle_t\langle n_j\rangle_t=\delta_{ij}D_{ii}(t)-D_{ij}(t)D_{ji}(t)$.
\begin{figure}[!htb]
	\centering
	\includegraphics[width=8.5cm]{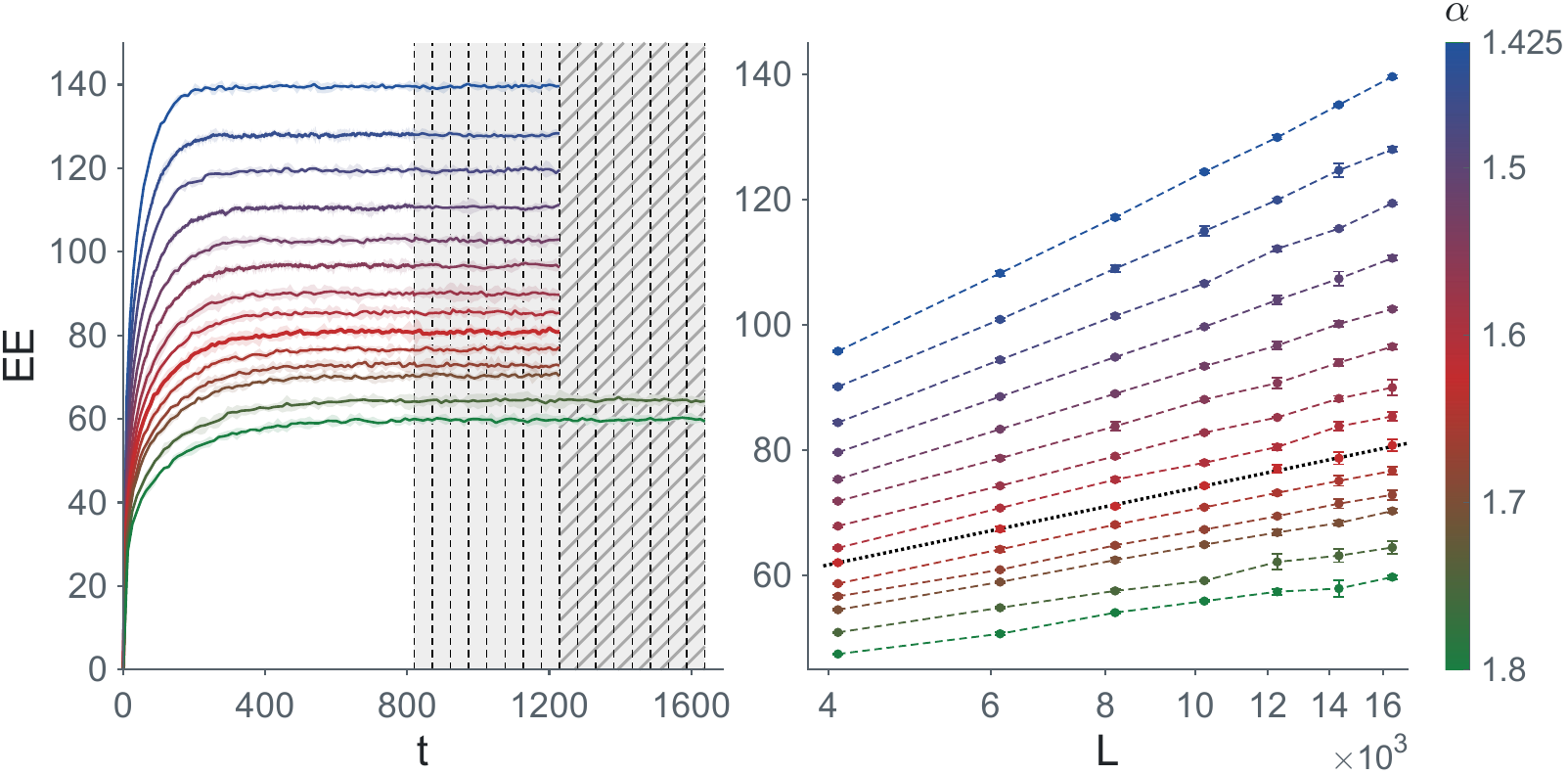}
	\caption{Entanglement entropy (EE) at $\gamma=0.5$ for hopping exponents $\alpha\in[1.425,1.8]$ and system sizes $L\in[4096,16384]$. The lightly shaded bands (left) and error bars (right) indicate two-sided $95\%$ Student-$t$ confidence intervals, computed by pooling individual trajectories across all disorder realizations. \textbf{(Left)} Time evolution of EE along quantum trajectories at $L=16384$ up to the scrambling stage. Dotted black lines mark the nine equally spaced sampling times in each window. Plain shading indicates $[\tau/4,3\tau/8]$ for $\alpha\leq 1.7$, and diagonal hatching indicates $[3\tau/8,\tau/2]$ for $\alpha>1.7$. \textbf{(Right)} EE as a function of $L$, averaged first over the sampled times within each trajectory and then over the ensemble of trajectories across disorder realizations. The black dotted line shows a logarithmic fit ($S\sim\log L$) to the data at the point $\alpha=1.625$, near the transition.
        }\label{fig:monitorEEL}
\end{figure}

\begin{figure}[!htb]
	\centering
	\includegraphics[width=8.5cm]{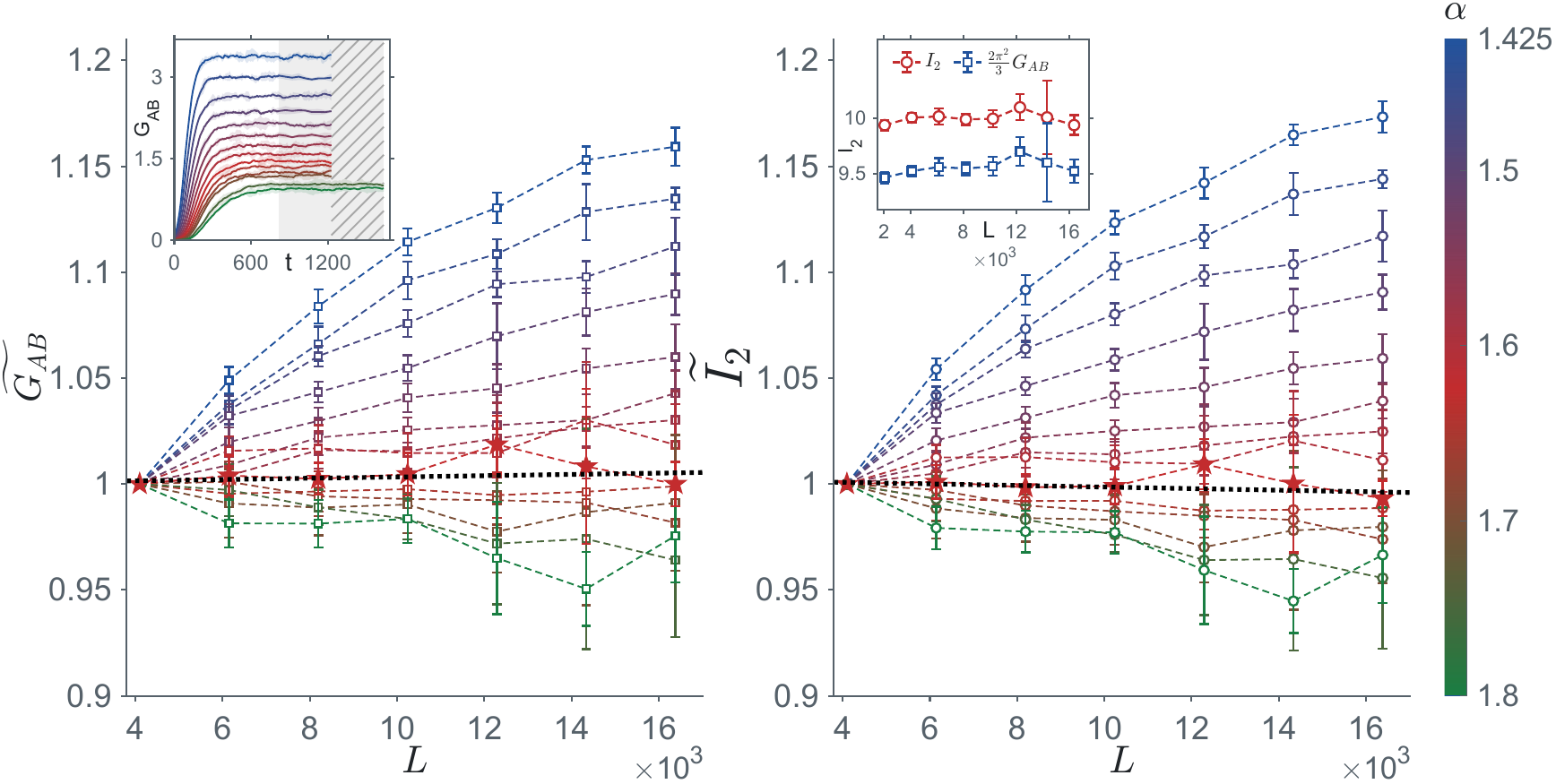}
	\caption{Magnitude of the particle-number covariance $G_{AB}$ (\textbf{left}) and mutual information $I_2$ (\textbf{right}) at $\gamma=0.5$ for hopping exponents $\alpha\in[1.425,1.8]$ and system sizes $L\in[4096,16384]$. Error bars indicate two-sided $95\%$ Student-$t$ confidence intervals, computed by pooling individual trajectories across all disorder realizations. For each $\alpha$, both observables are normalized by their respective values at $L=4096$: $\widetilde{G_{AB}}=G_{AB}/G_{AB}(L=4096)$ and $\widetilde{I_2}=I_2/I_2(L=4096)$. Red stars mark the data at $\alpha=1.625$, near the transition. The nearly flat black dotted lines show linear fits to these data. \textbf{(Left inset)} Time evolution of $G_{AB}$ along quantum trajectories at $L=16384$ up to the scrambling stage, using the same sampling windows as in the left panel of Fig.~\ref{fig:monitorEEL}. \textbf{(Right inset)} Comparison of $I_2$ and $\frac{2\pi^2}{3}G_{AB}$ at $\alpha=1.625$ for sizes from $L=2048$ to $16384$. The difference remains small across the sizes shown.  }\label{fig:monitorI2GAB}
\end{figure}

In Fig.~\ref{fig:monitorEEL}, we plot the entanglement entropy (EE) under continuous monitoring, whose time evolution exhibits an initial growth regime followed by a long-time plateau and is characterized by the timescale $\tau\equiv\frac{L}{b}$, with the scrambling time proportional to $\tau$ and decreasing as the measurement rate $\gamma$ and hopping exponent $\alpha$ increase.
Once both EE and $I_2/G_{AB}$ have reached their plateaus, as also shown in the inset of Fig.~\ref{fig:monitorI2GAB}(left), we compute the observables by averaging over nine equally spaced time points. For $\gamma=0.5,0.7$, we choose the averaging intervals $[\frac{\tau}{4},\frac{3\tau}{8}]$ for $\alpha\leq 1.7$ and $[\frac{3\tau}{8},\frac{\tau}{2}]$ for $\alpha>1.7$, while for $\gamma=0.3$, the corresponding intervals are $[\frac{3\tau}{8},\frac{\tau}{2}]$ and $[\frac{\tau}{2},\frac{5\tau}{8}]$, respectively. Near the critical point $\alpha_c$, as shown for $\alpha=1.625$ in Fig.~\ref{fig:monitorI2GAB}(right), we find the expected logarithmic scaling, while the same fitting form remains valid over a wide range of $\alpha$ near the transition.

For two disjoint blocks $A=\{1,\ldots,L/4\}$ and $B=\{L/2+1,\ldots,3L/4\}$, each of length $L/4$ and separated by gaps of length $L/4$ in both directions around the ring, we define their mutual information as
\begin{equation}\label{eq:I2}
I_2(t)=S_A(t)+S_B(t)-S_{A\cup B}(t).
\end{equation}
The mutual information cancels the local boundary contributions to EE and thus approaches a constant at criticality, providing a robust way to locate the transition point through scale-invariant crossings in finite-size scaling analysis. At leading order in the cumulant expansion, $I_2\boldsymbol{\approx}\frac{2\pi^2}{3}\overline{G_{AB}}$, where $G_{AB}=-\sum\limits_{i\in A,j\in B} C_{ij}(t)=\sum\limits_{i\in A,j\in B}|D_{ij}(t)|^2$ is the negative particle-number covariance between the two blocks.

In Fig.~\ref{fig:monitorI2GAB}, we plot the scaling of both $I_2$ and $G_{AB}$, normalizing each by its value at $L=4096$ to highlight their strong similarity in scaling behavior. For both observables, at $\alpha_c=1.625$, the absolute values of $I_2$ and $G_{AB}$ show no system-size dependence, confirming the scale-invariant behavior expected at criticality and supporting the use of universal crossings to locate the transition in the main text. We also compare the values of $I_2$ and $\frac{2\pi^2}{3}G_{AB}$ for $L\in[2048,16384]$, as shown in the inset of Fig.~\ref{fig:monitorI2GAB}(right), and find that their overall magnitudes differ by less than $5\%$, with no systematic system-size dependence of this difference, supporting the leading-order cumulant approximation while revealing a clear subleading contribution.

\begin{figure}[!htb]
	\centering
	\includegraphics[width=8.5cm]{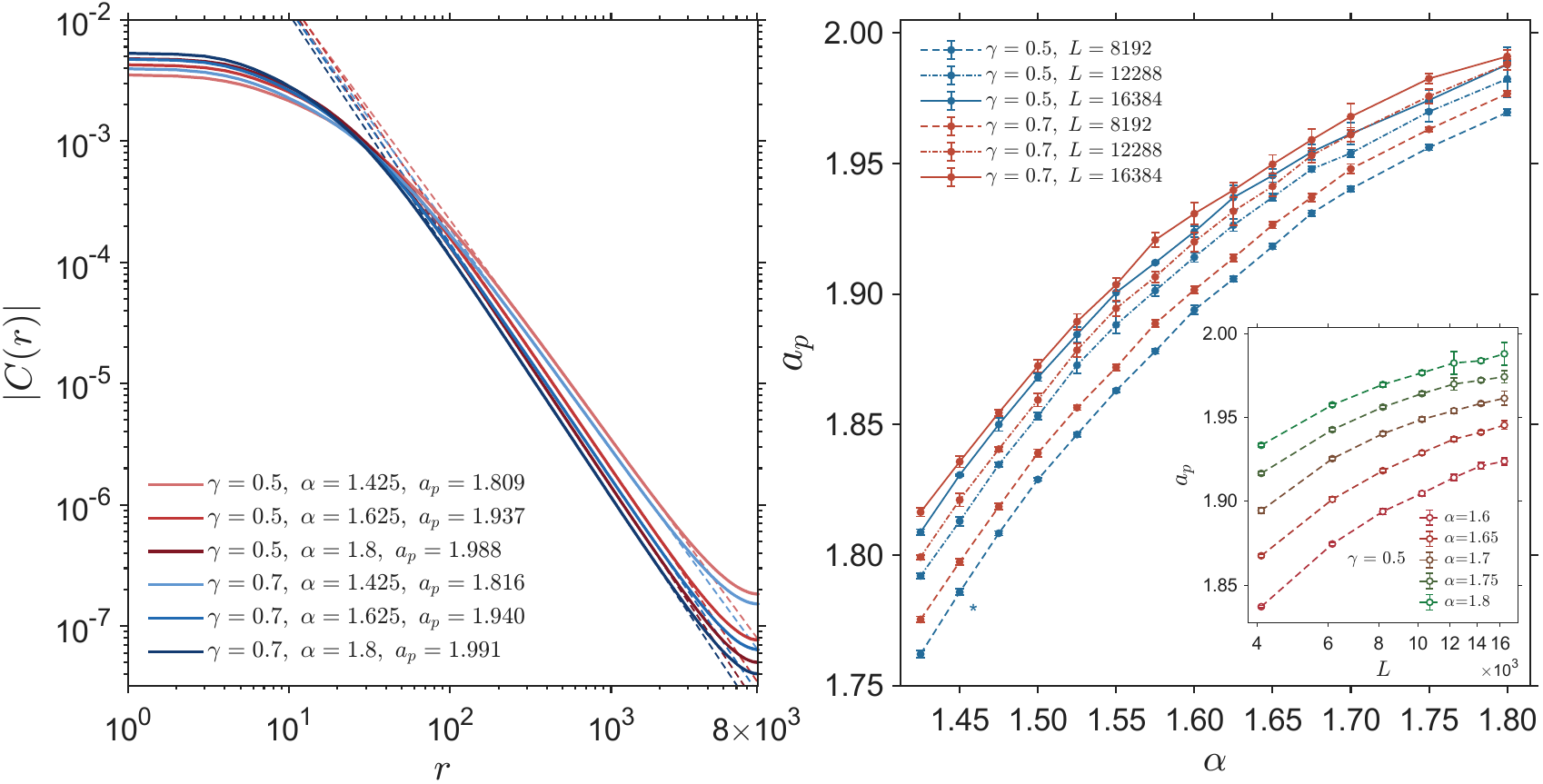}
	\caption{Density--density correlation function $C(r)$ and its effective power-law decay exponent $a_p$ at monitoring rates $\gamma=0.5$ and $0.7$. \textbf{(Left)} Log--log plot of the absolute value $|C(r)|$ at $L=16384$ for $\alpha=1.425,1.625,1.8$. Error bars are omitted because they are generally narrower than the plotted lines at the scale shown. The solid curves show geometric averages taken first over sampled times and then over trajectories across disorder realizations. The dashed lines show power-law fits over selected tail intervals. \textbf{(Right)} $a_p$ as a function of $\alpha$ for $\gamma=0.5$ and $0.7$ at $L=8192,12288,16384$. Error bars indicate two-sided $95\%$ Student-$t$ confidence intervals, computed by pooling individual trajectories across all disorder realizations. For each trajectory, the decay exponent is fitted to its time-averaged $C(r)$, and the resulting exponents are averaged geometrically. At fixed $L$, $a_p$ increases with both $\gamma$ and $\alpha$. Two finite-size effects are evident. First, $a_p$ increases with $L$ at fixed $\gamma$ and $\alpha$. Second, the $a_p(\alpha)$ curves flatten above the transition ($\alpha>\alpha_c$). The inset shows the slower-than-logarithmic growth of $a_p(L)$ at $\gamma=0.5$ for $L\in[4096,16384]$ (semilogx). We expect $a_p$ to continue increasing beyond $L=16384$ and eventually saturate at a larger but finite system size.}\label{fig:monitorCrap}
\end{figure}

The metal-insulator transition is clearly reflected in the power-law decay exponent $a_p$ of the density-density correlation function $C(r)$. At criticality, $C(r)\sim r^{-2}$ and $a_p=2$, with $a_p$ expected to decrease below the transition and increase above it. However, for $\alpha>\frac{3}{2}$, the superdiffusive kernel becomes diffusive, giving $C(r)\sim r^{-2}$ at short distances $r\ll\xi$, where $\xi$ is the localization length, so that at insufficiently large system sizes, the fitted exponent bends toward $a_p\approx 2$, as shown in Fig.~\ref{fig:monitorCrap}, explaining the shift from the theoretically expected value $\alpha_c=\frac32$ to the larger value $\alpha_c\approx 1.6$.  The size dependence of $a_p$ is evident in the inset of Fig.~\ref{fig:monitorCrap}(left) for $\gamma=0.5$, where at small $\alpha$, the shift in $a_p$ between $L=4096$ and $16384$ exceeds $0.1$, highlighting the need for large system sizes $L\in[8192,16384]$, over which the shift falls below $0.05$.

\acknowledgments
 {\it Acknowledgments.-} C.Y. thanks Bo Fan for pointing out the numerical transition point at $\alpha_c\approx 2$ in the quench dynamics of a Néel state.
C.Y. thanks Antonio Miguel Garcia-Garcia and Bo Fan for illuminating discussions about the numerical analysis. 
C.Y. acknowledges support from a T.D. Lee scholarship.

\bibliography{PRBM_reference.bib}

@article{MirlinEtAl1996,
  author        = {Mirlin, Alexander D. and Fyodorov, Yan V. and Dittes, Frank-Michael and Quezada, Javier and Seligman, Thomas H.},
  title         = {Transition from localized to extended eigenstates in the ensemble of power-law random banded matrices},
  journal       = {Phys. Rev. E},
  volume        = {54},
  number        = {4},
  pages         = {3221--3230},
  year          = {1996},
  doi           = {10.1103/PhysRevE.54.3221},
  url           = {https://doi.org/10.1103/PhysRevE.54.3221},
  eprint        = {cond-mat/9604163},
  archivePrefix = {arXiv}
}

@article{Peschel2003,
  author        = {Peschel, Ingo},
  title         = {Calculation of reduced density matrices from correlation functions},
  journal       = {J. Phys. A: Math. Gen.},
  volume        = {36},
  number        = {14},
  pages         = {L205--L208},
  year          = {2003},
  doi           = {10.1088/0305-4470/36/14/101},
  url           = {https://doi.org/10.1088/0305-4470/36/14/101},
  eprint        = {cond-mat/0212631},
  archivePrefix = {arXiv}
}

@article{EversMirlin2008,
  author        = {Evers, Ferdinand and Mirlin, Alexander D.},
  title         = {{Anderson} transitions},
  journal       = {Rev. Mod. Phys.},
  volume        = {80},
  number        = {4},
  pages         = {1355--1417},
  year          = {2008},
  doi           = {10.1103/RevModPhys.80.1355},
  url           = {https://doi.org/10.1103/RevModPhys.80.1355},
  eprint        = {0707.4378},
  archivePrefix = {arXiv},
  primaryClass  = {cond-mat.mes-hall}
}

@article{Anderson1958,
  author        = {Anderson, P. W.},
  title         = {Absence of Diffusion in Certain Random Lattices},
  journal       = {Phys. Rev.},
  volume        = {109},
  number        = {5},
  pages         = {1492--1505},
  year          = {1958},
  doi           = {10.1103/PhysRev.109.1492},
  url           = {https://doi.org/10.1103/PhysRev.109.1492}
}

@article{AbrahamsEtAl1979,
  author        = {Abrahams, E. and Anderson, P. W. and Licciardello, D. C. and Ramakrishnan, T. V.},
  title         = {Scaling Theory of Localization: Absence of Quantum Diffusion in Two Dimensions},
  journal       = {Phys. Rev. Lett.},
  volume        = {42},
  number        = {10},
  pages         = {673--676},
  year          = {1979},
  doi           = {10.1103/PhysRevLett.42.673},
  url           = {https://doi.org/10.1103/PhysRevLett.42.673}
}

@article{MirlinEvers2000,
  author        = {Mirlin, A. D. and Evers, F.},
  title         = {Multifractality and critical fluctuations at the {Anderson} transition},
  journal       = {Phys. Rev. B},
  volume        = {62},
  number        = {12},
  pages         = {7920--7933},
  year          = {2000},
  doi           = {10.1103/PhysRevB.62.7920},
  url           = {https://doi.org/10.1103/PhysRevB.62.7920},
  eprint        = {cond-mat/0003332},
  archivePrefix = {arXiv}
}

@article{GarciaGarciaWang2005,
  author        = {Garc{\'{\i}}a-Garc{\'{\i}}a, Antonio M. and Wang, Jiao},
  title         = {{Anderson} Transition in Quantum Chaos},
  journal       = {Phys. Rev. Lett.},
  volume        = {94},
  number        = {24},
  pages         = {244102},
  year          = {2005},
  doi           = {10.1103/PhysRevLett.94.244102},
  url           = {https://doi.org/10.1103/PhysRevLett.94.244102},
  eprint        = {cond-mat/0412367},
  archivePrefix = {arXiv},
  primaryClass  = {cond-mat.dis-nn}
}

@article{Mott1967,
  author        = {Mott, N. F.},
  title         = {Electrons in disordered structures},
  journal       = {Adv. Phys.},
  volume        = {16},
  number        = {61},
  pages         = {49--144},
  year          = {1967},
  doi           = {10.1080/00018736700101265},
  url           = {https://doi.org/10.1080/00018736700101265}
}

@article{ShklovskiiEtAl1993,
  author        = {Shklovskii, B. I. and Shapiro, B. and Sears, B. R. and Lambrianides, P. and Shore, H. B.},
  title         = {Statistics of spectra of disordered systems near the metal-insulator transition},
  journal       = {Phys. Rev. B},
  volume        = {47},
  number        = {17},
  pages         = {11487--11490},
  year          = {1993},
  doi           = {10.1103/PhysRevB.47.11487},
  url           = {https://doi.org/10.1103/PhysRevB.47.11487}
}

@article{AbaninEtAl2019,
  author        = {Abanin, Dmitry A. and Altman, Ehud and Bloch, Immanuel and Serbyn, Maksym},
  title         = {Colloquium: Many-body localization, thermalization, and entanglement},
  journal       = {Rev. Mod. Phys.},
  volume        = {91},
  number        = {2},
  pages         = {021001},
  year          = {2019},
  doi           = {10.1103/RevModPhys.91.021001},
  url           = {https://doi.org/10.1103/RevModPhys.91.021001},
  eprint        = {1804.11065},
  archivePrefix = {arXiv},
  primaryClass  = {cond-mat.dis-nn}
}

@article{AltshulerEtAl1997,
  author        = {Altshuler, Boris L. and Gefen, Yuval and Kamenev, Alex and Levitov, Leonid S.},
  title         = {Quasiparticle Lifetime in a Finite System: A Nonperturbative Approach},
  journal       = {Phys. Rev. Lett.},
  volume        = {78},
  number        = {14},
  pages         = {2803--2806},
  year          = {1997},
  doi           = {10.1103/PhysRevLett.78.2803},
  url           = {https://doi.org/10.1103/PhysRevLett.78.2803},
  eprint        = {cond-mat/9609132},
  archivePrefix = {arXiv}
}

@article{FyodorovMirlin1991,
  author        = {Fyodorov, Yan V. and Mirlin, Alexander D.},
  title         = {Scaling properties of localization in random band matrices: A {$\sigma$}-model approach},
  journal       = {Phys. Rev. Lett.},
  volume        = {67},
  number        = {18},
  pages         = {2405--2409},
  year          = {1991},
  doi           = {10.1103/PhysRevLett.67.2405},
  url           = {https://doi.org/10.1103/PhysRevLett.67.2405}
}

@article{Efetov1983,
  author        = {Efetov, K. B.},
  title         = {Supersymmetry and theory of disordered metals},
  journal       = {Adv. Phys.},
  volume        = {32},
  number        = {1},
  pages         = {53--127},
  year          = {1983},
  doi           = {10.1080/00018738300101531},
  url           = {https://doi.org/10.1080/00018738300101531}
}

@article{KamenevLevchenko2009,
  author        = {Kamenev, Alex and Levchenko, Alex},
  title         = {{Keldysh} technique and non-linear {$\sigma$}-model: basic principles and applications},
  journal       = {Adv. Phys.},
  volume        = {58},
  number        = {3},
  pages         = {197--319},
  year          = {2009},
  doi           = {10.1080/00018730902850504},
  url           = {https://doi.org/10.1080/00018730902850504},
  eprint        = {0901.3586},
  archivePrefix = {arXiv},
  primaryClass  = {cond-mat.other}
}

@article{GioevKlich2006,
  author        = {Gioev, Dimitri and Klich, Israel},
  title         = {Entanglement entropy of fermions in any dimension and the {Widom} conjecture},
  journal       = {Phys. Rev. Lett.},
  volume        = {96},
  pages         = {100503},
  year          = {2006},
  doi           = {10.1103/PhysRevLett.96.100503},
  url           = {https://doi.org/10.1103/PhysRevLett.96.100503}
}

@article{Daley2014,
  author        = {Daley, Andrew J.},
  title         = {Quantum trajectories and open many-body quantum systems},
  journal       = {Adv. Phys.},
  volume        = {63},
  number        = {2},
  pages         = {77--149},
  year          = {2014},
  doi           = {10.1080/00018732.2014.933502},
  url           = {https://doi.org/10.1080/00018732.2014.933502}
}

@article{Noel2022,
  author        = {Noel, Crystal and Niroula, Pradeep and Zhu, Daiwei and Risinger, Andrew and Egan, Laird and Biswas, Debopriyo and Cetina, Marko and Gorshkov, Alexey V. and Gullans, Michael J. and Huse, David A. and Monroe, Christopher},
  title         = {Measurement-induced quantum phases realized in a trapped-ion quantum computer},
  journal       = {Nat. Phys.},
  volume        = {18},
  number        = {7},
  pages         = {760--764},
  year          = {2022},
  doi           = {10.1038/s41567-022-01619-7},
  url           = {https://doi.org/10.1038/s41567-022-01619-7}
}

@article{Koh2023,
  author        = {Koh, Jin Ming and Sun, Shi-Ning and Motta, Mario and Minnich, Austin J.},
  title         = {Measurement-induced entanglement phase transition on a superconducting quantum processor with mid-circuit readout},
  journal       = {Nat. Phys.},
  volume        = {19},
  number        = {9},
  pages         = {1314--1319},
  year          = {2023},
  doi           = {10.1038/s41567-023-02076-6},
  url           = {https://doi.org/10.1038/s41567-023-02076-6}
}

@article{Bergquist1986,
  author        = {Bergquist, J. C. and Hulet, Randall G. and Itano, Wayne M. and Wineland, D. J.},
  title         = {Observation of Quantum Jumps in a Single Atom},
  journal       = {Phys. Rev. Lett.},
  volume        = {57},
  number        = {14},
  pages         = {1699--1702},
  year          = {1986},
  doi           = {10.1103/physrevlett.57.1699},
  url           = {https://doi.org/10.1103/physrevlett.57.1699}
}

@article{Murch2013,
  author        = {Murch, K. W. and Weber, S. J. and Macklin, C. and Siddiqi, I.},
  title         = {Observing single quantum trajectories of a superconducting quantum bit},
  journal       = {Nature},
  volume        = {502},
  number        = {7470},
  pages         = {211--214},
  year          = {2013},
  doi           = {10.1038/nature12539},
  url           = {https://doi.org/10.1038/nature12539}
}

@article{WisemanMilburn1993,
  author        = {Wiseman, H. M. and Milburn, G. J.},
  title         = {Quantum theory of optical feedback via homodyne detection},
  journal       = {Phys. Rev. Lett.},
  volume        = {70},
  number        = {5},
  pages         = {548--551},
  year          = {1993},
  doi           = {10.1103/PhysRevLett.70.548},
  url           = {https://doi.org/10.1103/PhysRevLett.70.548}
}

@article{Minev2019,
  author        = {Minev, Z. K. and Mundhada, S. O. and Shankar, S. and Reinhold, P. and Guti{\'e}rrez-Jáuregui, R. and Schoelkopf, R. J. and Mirrahimi, M. and Carmichael, H. J. and Devoret, M. H.},
  title         = {To catch and reverse a quantum jump mid-flight},
  journal       = {Nature},
  volume        = {570},
  number        = {7760},
  pages         = {200--204},
  year          = {2019},
  doi           = {10.1038/s41586-019-1287-z},
  url           = {https://doi.org/10.1038/s41586-019-1287-z}
}

@article{Sayrin2011,
  author        = {Sayrin, Cl{\'e}ment and Dotsenko, Igor and Zhou, Xingxing and Peaudecerf, Bruno and Rybarczyk, Th{\'e}o and Gleyzes, S{\'e}bastien and Rouchon, Pierre and Mirrahimi, Mazyar and Amini, Hadis and Brune, Michel and Raimond, Jean-Michel and Haroche, Serge},
  title         = {Real-time quantum feedback prepares and stabilizes photon number states},
  journal       = {Nature},
  volume        = {477},
  number        = {7362},
  pages         = {73--77},
  year          = {2011},
  doi           = {10.1038/nature10376},
  url           = {https://doi.org/10.1038/nature10376}
}

@article{LeGal2024,
  author        = {Le Gal, Youenn and Turkeshi, Xhek and Schir{\`o}, Marco},
  title         = {Entanglement Dynamics in Monitored Systems and the Role of Quantum Jumps},
  journal       = {PRX Quantum},
  volume        = {5},
  number        = {3},
  pages         = {030329},
  year          = {2024},
  doi           = {10.1103/prxquantum.5.030329},
  url           = {https://doi.org/10.1103/prxquantum.5.030329}
}

@article{Fan2026SingleMeasurement,
  author        = {Fan, Bo and Yin, Can and Garc{\'{\i}}a-Garc{\'{\i}}a, Antonio M.},
  title         = {Destruction and recovery of the entanglement entropy of a many-body quantum system after a single measurement},
  journal       = {Phys. Rev. B},
  volume        = {113},
  number        = {5},
  pages         = {054304},
  year          = {2026},
  doi           = {10.1103/qwtt-66nh},
  url           = {https://doi.org/10.1103/qwtt-66nh}
}

@article{Poboiko2023,
  author        = {Poboiko, Igor and P{\"o}pperl, Paul and Gornyi, Igor V. and Mirlin, Alexander D.},
  title         = {Theory of Free Fermions under Random Projective Measurements},
  journal       = {Phys. Rev. X},
  volume        = {13},
  number        = {4},
  pages         = {041046},
  year          = {2023},
  doi           = {10.1103/physrevx.13.041046},
  url           = {https://doi.org/10.1103/physrevx.13.041046}
}

@article{Fan2026GPU,
  author        = {Fan, Bo and Yin, Can and Garc{\'{\i}}a-Garc{\'{\i}}a, Antonio M.},
  title         = {Entanglement dynamics of monitored noninteracting fermions on graphics processing units},
  journal       = {Phys. Rev. B},
  volume        = {113},
  number        = {9},
  pages         = {094317},
  year          = {2026},
  doi           = {10.1103/x7vx-tqr4},
  url           = {https://doi.org/10.1103/x7vx-tqr4}
}

@article{Minato2022,
  author        = {Minato, Takaaki and Sugimoto, Koudai and Kuwahara, Tomotaka and Saito, Keiji},
  title         = {Fate of Measurement-Induced Phase Transition in Long-Range Interactions},
  journal       = {Phys. Rev. Lett.},
  volume        = {128},
  number        = {1},
  pages         = {010603},
  year          = {2022},
  doi           = {10.1103/PhysRevLett.128.010603},
  url           = {https://doi.org/10.1103/PhysRevLett.128.010603},
  eprint        = {2104.09118},
  archivePrefix = {arXiv}
}

@article{Muller2022,
  author        = {M{\"u}ller, Thomas and Diehl, Sebastian and Buchhold, Michael},
  title         = {Measurement-Induced Dark State Phase Transitions in Long-Ranged Fermion Systems},
  journal       = {Phys. Rev. Lett.},
  volume        = {128},
  number        = {1},
  pages         = {010605},
  year          = {2022},
  doi           = {10.1103/PhysRevLett.128.010605},
  url           = {https://doi.org/10.1103/PhysRevLett.128.010605},
  eprint        = {2105.08076},
  archivePrefix = {arXiv}
}

@misc{Fan2026LongRange,
  author        = {Fan, Bo and Garc{\'{\i}}a-Garc{\'{\i}}a, Antonio M.},
  title         = {Entanglement transitions and multifractality in monitored free-fermions with random long-range hopping},
  year          = {2026},
  eprint        = {2607.23871},
  archivePrefix = {arXiv},
  url           = {https://arxiv.org/abs/2607.23871}
}

@article{LiChenFisher2018,
  author  = {Li, Yaodong and Chen, Xiao and Fisher, Matthew P. A.},
  title   = {Quantum Zeno effect and the many-body entanglement transition},
  journal = {Phys. Rev. B},
  volume  = {98},
  pages   = {205136},
  year    = {2018},
  doi     = {10.1103/PhysRevB.98.205136},
  url     = {https://doi.org/10.1103/PhysRevB.98.205136}
}

@article{SkinnerRuhmanNahum2019,
  author  = {Skinner, Brian and Ruhman, Jonathan and Nahum, Adam},
  title   = {Measurement-Induced Phase Transitions in the Dynamics of Entanglement},
  journal = {Phys. Rev. X},
  volume  = {9},
  pages   = {031009},
  year    = {2019},
  doi     = {10.1103/PhysRevX.9.031009},
  url     = {https://doi.org/10.1103/PhysRevX.9.031009}
}

@article{Hoke2023,
  author  = {{Google Quantum AI and Collaborators}},
  title   = {Measurement-induced entanglement and teleportation on a noisy quantum processor},
  journal = {Nature},
  volume  = {622},
  pages   = {481--486},
  year    = {2023},
  doi     = {10.1038/s41586-023-06505-7},
  url     = {https://doi.org/10.1038/s41586-023-06505-7}
}

@article{Gurvitz2000,
  author  = {Gurvitz, S. A.},
  title   = {Delocalization in the {Anderson} Model due to a Local Measurement},
  journal = {Phys. Rev. Lett.},
  volume  = {85},
  pages   = {812--815},
  year    = {2000},
  doi     = {10.1103/PhysRevLett.85.812},
  url     = {https://doi.org/10.1103/PhysRevLett.85.812}
}

@article{YiEtAl2011,
  author  = {Yi, Juyeon and Talkner, Peter and Ingold, Gert-Ludwig},
  title   = {Approaching infinite temperature upon repeated measurements of a quantum system},
  journal = {Phys. Rev. A},
  volume  = {84},
  pages   = {032121},
  year    = {2011},
  doi     = {10.1103/PhysRevA.84.032121},
  url     = {https://doi.org/10.1103/PhysRevA.84.032121}
}

@article{BuchholdEtAl2021,
  author  = {Buchhold, M. and Minoguchi, Y. and Altland, A. and Diehl, S.},
  title   = {Effective Theory for the Measurement-Induced Phase Transition of {Dirac} Fermions},
  journal = {Phys. Rev. X},
  volume  = {11},
  pages   = {041004},
  year    = {2021},
  doi     = {10.1103/PhysRevX.11.041004},
  url     = {https://doi.org/10.1103/PhysRevX.11.041004}
}

@article{GuoEtAl2025,
  author  = {Guo, Haoyu and Foster, Matthew S. and Jian, Chao-Ming and Ludwig, Andreas W. W.},
  title   = {Field theory of monitored interacting fermion dynamics with charge conservation},
  journal = {Phys. Rev. B},
  volume  = {112},
  pages   = {064304},
  year    = {2025},
  doi     = {10.1103/gdxd-pw8v},
  url     = {https://doi.org/10.1103/gdxd-pw8v},
  eprint  = {2410.07317},
  archivePrefix = {arXiv}
}

@article{Poboiko2025Interacting,
  author        = {Poboiko, Igor and P{\"o}pperl, Paul and Gornyi, Igor V. and Mirlin, Alexander D.},
  title         = {Measurement-induced transitions for interacting fermions},
  journal       = {Phys. Rev. B},
  volume        = {111},
  pages         = {024204},
  year          = {2025},
  doi           = {10.1103/PhysRevB.111.024204},
  url           = {https://doi.org/10.1103/PhysRevB.111.024204},
  eprint        = {2410.07334},
  archivePrefix = {arXiv}
}

@misc{YinEtAl2026Disorder,
  author        = {Yin, Can and Fan, Bo and Garc{\'{\i}}a-Garc{\'{\i}}a, Antonio M.},
  title         = {No measurement induced phase transition in the entanglement dynamics of monitored non-interacting one-dimensional fermions in a disordered or quasiperiodic potential},
  year          = {2026},
  eprint        = {2605.10758},
  archivePrefix = {arXiv},
  primaryClass  = {quant-ph},
  url           = {https://arxiv.org/abs/2605.10758}
}

@article{SkinnerEtAl2019,
  title = {Measurement-Induced Phase Transitions in the Dynamics of Entanglement},
  author = {Skinner, Brian and Ruhman, Jonathan and Nahum, Adam},
  journal = {Phys. Rev. X},
  volume = {9},
  pages = {031009},
  year = {2019},
  doi = {10.1103/PhysRevX.9.031009}
}

@article{VasseurEtAl2019,
  title = {Entanglement Transitions from Holographic Random Tensor Networks},
  author = {Vasseur, Romain and Potter, Andrew C. and You, Yi-Zhuang and Ludwig, Andreas W. W.},
  journal = {Phys. Rev. B},
  volume = {100},
  pages = {134203},
  year = {2019},
  doi = {10.1103/PhysRevB.100.134203}
}

@article{JianEtAl2020,
  title = {Measurement-Induced Criticality in Random Quantum Circuits},
  author = {Jian, Chao-Ming and You, Yi-Zhuang and Vasseur, Romain and Ludwig, Andreas W. W.},
  journal = {Phys. Rev. B},
  volume = {101},
  pages = {104302},
  year = {2020},
  doi = {10.1103/PhysRevB.101.104302}
}

@article{JianEtAl2022,
  title = {Criticality and Entanglement in Nonunitary Quantum Circuits and Tensor Networks of Noninteracting Fermions},
  author = {Jian, Chao-Ming and Bauer, Bela and Keselman, Anna and Ludwig, Andreas W. W.},
  journal = {Phys. Rev. B},
  volume = {106},
  pages = {134206},
  year = {2022},
  doi = {10.1103/PhysRevB.106.134206}
}

@article{FavaEtAl2023,
  title = {Nonlinear Sigma Models for Monitored Dynamics of Free Fermions},
  author = {Fava, Michele and Piroli, Lorenzo and Swann, Tobias and Bernard, Denis and Nahum, Adam},
  journal = {Phys. Rev. X},
  volume = {13},
  pages = {041045},
  year = {2023},
  doi = {10.1103/PhysRevX.13.041045}
}

@article{NahumWiese2023,
  title = {Renormalization Group for Measurement and Entanglement Phase Transitions},
  author = {Nahum, Adam and Wiese, Kay J{\"o}rg},
  journal = {Phys. Rev. B},
  volume = {108},
  pages = {104203},
  year = {2023},
  doi = {10.1103/PhysRevB.108.104203}
}

@article{FavaEtAl2024,
  title = {Monitored Fermions with Conserved {$U(1)$} Charge},
  author = {Fava, Michele and Piroli, Lorenzo and Bernard, Denis and Nahum, Adam},
  journal = {Phys. Rev. Research},
  volume = {6},
  pages = {043246},
  year = {2024},
  doi = {10.1103/PhysRevResearch.6.043246}
}

@article{LiEtAl2021,
  title = {Conformal Invariance and Quantum Nonlocality in Critical Hybrid Circuits},
  author = {Li, Yaodong and Chen, Xiao and Ludwig, Andreas W. W. and Fisher, Matthew P. A.},
  journal = {Phys. Rev. B},
  volume = {104},
  pages = {104305},
  year = {2021},
  doi = {10.1103/PhysRevB.104.104305}
}

@article{PoboikoEtAl2024,
  title = {Measurement-Induced Phase Transition for Free Fermions above One Dimension},
  author = {Poboiko, Igor and Gornyi, Igor V. and Mirlin, Alexander D.},
  journal = {Phys. Rev. Lett.},
  volume = {132},
  pages = {110403},
  year = {2024},
  doi = {10.1103/PhysRevLett.132.110403}
}

@article{Levitov1990,
  title = {Delocalization of vibrational modes caused by electric dipole interaction},
  author = {Levitov, L. S.},
  journal = {Phys. Rev. Lett.},
  volume = {64},
  issue = {5},
  pages = {547--550},
  numpages = {0},
  year = {1990},
  month = {Jan},
  publisher = {American Physical Society},
  doi = {10.1103/PhysRevLett.64.547},
  url = {https://link.aps.org/doi/10.1103/PhysRevLett.64.547}
}

@article{PouranvariYang2014,
  title = {Maximally entangled mode, metal-insulator transition, and violation of entanglement area law in noninteracting fermion ground states},
  author = {Pouranvari, Mohammad and Yang, Kun},
  journal = {Phys. Rev. B},
  volume = {89},
  pages = {115104},
  year = {2014},
  month = {Mar},
  publisher = {American Physical Society},
  doi = {10.1103/PhysRevB.89.115104},
  url = {https://doi.org/10.1103/PhysRevB.89.115104}
}

@article{KlichLevitov2009,
  title   = {Quantum Noise as an Entanglement Meter},
  author  = {Klich, Israel and Levitov, Leonid},
  journal = {Phys. Rev. Lett.},
  volume  = {102},
  pages   = {100502},
  year    = {2009},
  doi     = {10.1103/PhysRevLett.102.100502},
  eprint  = {0804.1377},
  archivePrefix = {arXiv},
  primaryClass  = {quant-ph}
}

@article{Burmistrov2017,
  title   = {Entanglement entropy and particle number cumulants of disordered fermions},
  author  = {Burmistrov, I. S. and Tikhonov, K. S. and Gornyi, I. V. and Mirlin, A. D.},
  journal = {Annals of Physics},
  volume  = {383},
  pages   = {140--156},
  year    = {2017},
  doi     = {10.1016/j.aop.2017.05.011},
  eprint  = {1701.05876},
  archivePrefix = {arXiv},
  primaryClass  = {cond-mat.mes-hall}
}

@article{SlevinOhtsuki2014,
  author  = {Keith Slevin and Tomi Ohtsuki},
  title   = {Critical exponent for the Anderson transition in the three-dimensional orthogonal universality class},
  journal = {New Journal of Physics},
  volume  = {16},
  pages   = {015012},
  year    = {2014},
  doi     = {10.1088/1367-2630/16/1/015012},
  eprint  = {1307.4483},
  archivePrefix = {arXiv}
}

@book{kamenev2011,
  author    = {Kamenev, Alex},
  title     = {Field Theory of Non-Equilibrium Systems},
  publisher = {Cambridge University Press},
  address   = {Cambridge},
  year      = {2011}
}

@unpublished{FanGarciaGarcia2026,
  author = {Fan, Bo and Garc{\'i}a-Garc{\'i}a, Antonio M.},
  note = {to be published}
}

\clearpage
\onecolumngrid
\setcounter{secnumdepth}{2}
\setcounter{section}{0}
\setcounter{subsection}{0}
\setcounter{equation}{0}
\setcounter{figure}{0}
\setcounter{table}{0}

\renewcommand{\thesection}{S\arabic{section}}
\renewcommand{\thesubsection}{S\arabic{section}.\arabic{subsection}}
\renewcommand{\theequation}{S\arabic{equation}}
\renewcommand{\thefigure}{S\arabic{figure}}
\renewcommand{\thetable}{S\arabic{table}}

\makeatletter
\renewcommand{\@hangfrom@section}[3]{\@hangfrom{#1#2}#3}
\makeatother

\begin{center}
{\large\bfseries Supplemental Materials\par}
\end{center}
This Supplemental Material presents analytical treatments of entanglement in the PRBM model in three settings: ground-state entanglement in equilibrium, long-time entanglement following a unitary quench from high-energy Slater-determinant initial states, and continuously monitored entanglement dynamics under the QSD protocol.

In Section 1, we analyze equilibrium ground-state entanglement by relating the second cumulant of particle-number fluctuations to the superdiffusive retarded-advanced (RA) ladder. This gives the entanglement transition at $\alpha_c=1$ and the critical scaling $S_A\sim(\log L)^2$.

In Section 2, we investigate long-time entanglement following a unitary quench from a high-energy Slater-determinant state. For $\alpha>\frac32$, a sparse two-level resonance analysis gives $C(r)\sim r^{-\alpha}$, the marginal entanglement scaling $S_A\sim\log L$ at $\alpha_c=2$, and $I_2\sim 8G_{AB}$ within the independent-resonance approximation.

In Section 3, we develop the replicated Keldysh formalism for continuously monitored dynamics under the QSD protocol and derive the superdiffusive $SU(R)$ nonlinear sigma model in the replicon sector. The tree-level RG identifies the marginal boundary at $\alpha_c=\frac32$, while the replicon propagator gives $C(r)\sim r^{-(\alpha+\frac12)}$ for $1<\alpha<\frac32$ and scale-invariant $G_{AB}$ and $I_2$ at the marginal boundary.
\section{Equilibrium ground-state entanglement with no monitoring}
We consider ground-state entanglement in the one-dimensional many-body PRBM model. The quadratic Hamiltonian ensures a Gaussian ground state, which is fully characterized by its equal-time fermionic correlation matrix.
To relate these correlations to the single-particle spectrum, we start from the equilibrium KMS condition $G^>(t)=-G^<(t+i\beta)$, where $G^>_{ij}(t)=i\langle c_i(t)c_j^\dagger(0)\rangle$ and $G^<_{ij}(t)=-i\langle c_j^\dagger(0)c_i(t)\rangle$. In Fourier space, this gives $G^>(\omega)=-e^{\beta \omega}G^<(\omega)$, and hence $G^<(\omega)=n_F(\omega)(G^<(\omega)-G^>(\omega))$, where $n_F(\omega)=\frac{1}{e^{\beta \omega}+1}$ is the Fermi-Dirac distribution at chemical potential $\mu=0$, as the disorder ensemble has exact spectral symmetry about the band center, chosen as zero energy. Using the relation $G^R(\omega)-G^A(\omega)=G^<(\omega)-G^>(\omega)$, where $G^R(t)=-i\theta(t)\langle \{c(t),c^\dagger(0)\}\rangle$ and $G^A(t)=i\theta(-t)\langle \{c(t),c^\dagger(0)\}\rangle$, we obtain $G^<(\omega)=n_F(\omega)(G^R(\omega)-G^A(\omega))$.

We next compute the connected density-density correlation function $C_{ij}\equiv \langle \delta n_i \delta n_j\rangle=\langle n_i n_j\rangle-\langle n_i\rangle\langle n_j\rangle$, and, since the state is Gaussian, Wick's theorem gives $C_{ij}=-\langle c_i^\dagger c_j\rangle\langle c_j^\dagger c_i\rangle$ for $i\neq j$. Using the spectral representation and the statistical translational invariance restored by disorder averaging under periodic boundary conditions, with $C_{ij}=C(r)$ for $r=|i-j|$, we obtain~\cite{Burmistrov2017}
\begin{equation}
\begin{aligned}\label{eq:SCr}
C(r)&=\frac{1}{L}\sum_i \overline{\langle c_i^\dag c_{i+r}\rangle \langle c_i c_{i+r}^\dag \rangle} =\int_{q,k,k',\omega,\omega'} e^{-iqr} \overline{A_{kk'}(\omega) A_{k'+q,k+q}(\omega')}n_F(\omega) (1-n_F(\omega'))\\
&\approx\int_{\omega,\omega'} n_F(\omega) (1-n_F(\omega'))\int_{q,k,k'} \frac{e^{-i q r}}{2\pi^2} \operatorname{Re}\left(\overline{G^A_{kk'}(\omega)G^R_{k'+q,k+q}(\omega')}\right).
\end{aligned}
\end{equation}
where $A_{kk'}(\omega)=[G^A_{kk'}(\omega)-G^R_{kk'}(\omega)]/(2\pi i)$, $q$ is the momentum conjugate to $r$, and the disorder average $\overline{\cdot}$ is taken over the full product to preserve the two-particle correlations and restore statistical translational invariance. We retain the diffusive $RA$ sector, whose soft pole gives the leading infrared contribution, while the $RR$ and $AA$ contributions are subleading at the Gaussian-ladder level, as can be verified by direct integration. The spatial sum is normalized by $1/L$, and the average filling fraction $n\in(0,1]$ is defined as $n=\int d\omega\,n_F(\omega)A(\omega)$.

In the low-temperature limit, the equilibrium state approaches the ground state, and the Fermi-Dirac distribution reduces to $n_F(\epsilon)=\Theta(-\epsilon)$ as $T\to0$. Only transitions from occupied to unoccupied states contribute, and introducing the energy transfer $\omega=\epsilon'-\epsilon>0$ and center energy $E=(\epsilon+\epsilon')/2$ restricts the energy integration to the Fermi window $E\in[-\omega/2,\omega/2]$, so that $E$ is also small when $\omega$ is taken to be small, yielding
\begin{equation}
C(q)\approx\frac{1}{4\pi}\int_0^\infty d\omega\int_{-\omega/2}^{\omega/2}dE\int_{k,k'}\operatorname{Re}\overline{G^A_{kk'}(E-\omega/2)G^R_{k'+q,k+q}(E+\omega/2)}
=\frac{1}{4\pi}\int_0^\infty d\omega\,\omega\operatorname{Re}\Gamma(q,\omega),
\end{equation}
where the disorder-averaged $RA$ ladder, integrated over the Fermi window, is defined as
$\Gamma(q,\omega)=\frac{1}{\omega}\int_{-\omega/2}^{\omega/2}dE\int_{k,k'}\overline{G^A_{kk'}(E-\omega/2)G^R_{k'+q,k+q}(E+\omega/2)}$.
We consider the second cumulant of particle-number fluctuations in a subsystem of length $L$, $\langle\langle N^2\rangle\rangle=\int_0^L dx\int_0^L dy\,C(x-y)$. The Fourier transform of the subsystem gives the form factor $4\sin^2(qL/2)/q^2$, yielding
\begin{equation}\label{eq:N2}
\langle\langle N^2\rangle\rangle=\int\frac{dq}{2\pi}\frac{4\sin^2(qL/2)}{q^2}C(q)\approx\frac{1}{\pi^2}\int_0^\infty d\omega\,\omega\int_0^\infty dq\,\frac{\sin^2(qL/2)}{q^2}\operatorname{Re}\Gamma(q,\omega).
\end{equation}

Using the full counting statistics relation, the entanglement entropy is approximated by $S_A\approx(\pi^2/3)\langle\langle N^2\rangle\rangle$~\cite{KlichLevitov2009} when higher-order cumulants are subleading in the infrared, so its leading scaling is governed by the diffusive $RA$ ladder $\Gamma(q,\omega)$.
To determine the diffusive ladder, we first define the variance of PRBM randomness, $W(r)\equiv \overline{|h_{i,i+r}|^2}$. The Born self-energy $\Sigma$ starts at second order since $\overline{h_{ij}}=0$, $\Sigma^{(2)}_{ij}=\sum_{kl}\overline{h_{ik}h_{lj}} (G_0)_{kl}(E)=\delta_{ij}\sum_k W(i-k) (G_0)_{kk}(E)$, where we retain the leading diagonal contraction. The self-consistent Born approximation replaces the bare Green's function with the disorder-averaged dressed propagator $g^{R,A}(E)\equiv\overline{G^{R,A}_{ii}(E)}_{\mathrm{SCBA}}$, yielding
\begin{equation}
\Sigma^{R,A}(E)=\sum_r W(r)g^{R,A}(E)=W_0g^{R,A}(E),\qquad
g^{R,A}(E)=\frac{1}{E\pm i0^+-W_0g^{R,A}(E)}.
\end{equation}
Here we have used statistical translational invariance to define $W_0\equiv\sum_r W(r)$, and the second relation follows from the SCBA Dyson equation $g=[G_0^{-1}-\Sigma]^{-1}$. The SCBA solution yields a semicircular density of states, as in random matrix theory. The full Green's function $G$ can be formally expanded around the SCBA dressed Green's function $g$ as $G=g+g(h-\Sigma)g+g(h-\Sigma)g(h-\Sigma)g+\cdots$, where $h=(h_{ij})$ is the random Hamiltonian matrix. In the large-bandwidth limit, single-particle scattering is distributed over $O(b)$ hopping channels. As a result, additional contractions involving repeated hopping bonds are suppressed by a factor $\frac{\sum_r W(r)^2}{(\sum_r W(r))^2}\sim b^{-1}$, motivating the leading ladder approximation $\Gamma\simeq\Gamma^{\mathrm{ladder}}$ in the large-$b$ weak-interference regime, consistent with the perturbative expansion underlying the effective NLSM~\cite{MirlinEtAl1996}. We obtain the Bethe–Salpeter equation by resumming the ladder rungs,
\begin{equation}\label{eq:BS}
\begin{aligned}
\Gamma^{\mathrm{ladder}}_{ij}
&=g^Rg^A\delta_{ij}
+(g^Rg^A)^2W(i-j)
+(g^Rg^A)^3\sum_kW(i-k)W(k-j)+\cdots\\
&=g^Rg^A\delta_{ij}
+g^Rg^A\sum_kW(i-k)\Gamma^{\mathrm{ladder}}_{kj},\\
&\rightarrow\Gamma^{\mathrm{ladder}}(q,E,\omega)
=\frac{1}{(g^Rg^A)^{-1}-W(q)}
=\frac{1}{W_0-W(q)+\frac{-\omega-2i0^+}{g^R-g^A}}.
\end{aligned}
\end{equation}
where $W(q)=\sum_r e^{iqr}W(r)$ is the Fourier transform of the variance, we define $g^R\equiv g^R(E+\omega/2)$ and $g^A\equiv g^A(E-\omega/2)$ for brevity, and we use the SCBA identity $(g^Rg^A)^{-1}=W_0+\frac{-\omega-2i0^+}{g^R-g^A}$. Then we define the SCBA density of states $\nu(E)=-\frac{1}{\pi}\operatorname{Im}g^R(E)$, and use the low-frequency relation $g^R(E+\omega/2)-g^A(E-\omega/2)=-2\pi i\nu(E)+O(\omega)$, so that
\begin{equation}\label{eq:kernel}
\Gamma\approx\Gamma^{\mathrm{ladder}}
\approx\frac{2\pi\nu(E)}{2\pi\nu(E)[W_0-W(q)]-i\omega}
\approx\frac{2\pi\nu(0)}{D_\alpha|q|^{2\alpha-1}-i\omega},
\qquad D_\alpha\equiv D=4\pi\nu(0)c_\alpha b^{2\alpha}.
\end{equation}
where we have used $W_0-W(q)=2\sum_{r=1}^\infty W(r)(1-\cos qr)\approx 2c_\alpha b^{2\alpha}|q|^{2\alpha-1}$ with $c_\alpha=\int_0^\infty dx\,\frac{1-\cos x}{x^{2\alpha}}$, using the off-diagonal variance $W(r)\approx(b/r)^{2\alpha}$ in the $J=1$ convention, for $1/2<\alpha<3/2$ and $|q|b\ll1$.
For $\alpha>3/2$, we recover the diffusive kernel $Dq^2-i\omega$ from $W_0-W(q)=2\sum_{r=1}^\infty W(r)(1-\cos qr)=q^2\sum_{r=1}^{\infty}r^2W(r)+o(q^2)$.
We also take $\nu(E)\approx\nu(0)$ within the low-frequency Fermi window $E\in[-\omega/2,\omega/2]$.

We have found that the leading ground-state entanglement scaling is governed by the RA ladder $\Gamma$, whose long-wavelength kernel yields the tree-level scaling of the dimensionless stiffness $g(L)\propto\nu(0)D_\alpha L^{2-2\alpha}$, giving the RG flow
\begin{equation}
\frac{dg}{dl}=(2-2\alpha)g,\qquad l=\log L.
\end{equation}
As a result, the transition occurs at $\alpha_c=1$, and for $\alpha\in(1/2,1)$, the positive coefficient drives the stiffness toward the weak-coupling extended metallic regime as the RG length scale $l$ increases, whereas for $1<\alpha<3/2$, the negative coefficient drives the stiffness toward the strong-coupling localized regime, with the localization length $\xi\sim b^{(2\alpha-1)/(2\alpha-2)}$, following the Anderson-localization analysis of single-particle PRBM~\cite{MirlinEtAl1996}.
Near the transition, we expand the tree-level RG solution as $g(L)/g(L_0)=\exp[(2-2\alpha)\log(L/L_0)]=1+(2-2\alpha)\log(L/L_0)+\cdots$, giving logarithmic finite-size scaling in contrast to the conventional $L^{1/\nu}$ scaling, so using standard polynomial fitting with scaling fields $\phi_i=u_i(w)L^{1/\nu}$, where $u_i(w)$ is expanded polynomially in the reduced hopping exponent $w=(\alpha-\alpha_c)/\alpha_c$, the fitted $\nu$ increases with system size.

Then we substitute the superdiffusive RA ladder $\Gamma$ from Eq.~\ref{eq:kernel} into the expression for the second-order particle-number cumulant $\langle\langle N^2\rangle\rangle$ in Eq.~\ref{eq:N2}.
We introduce a frequency cutoff $\Lambda_\omega\sim1/\tau$ to restrict the calculation to the low-frequency sector, with $\tau$ being the relaxation time, and the frequency integration gives $D|q|^{2\alpha-1}\log[\Lambda_\omega/(D|q|^{2\alpha-1})]$ for $D|q|^{2\alpha-1}\ll\Lambda_\omega$.
With the momentum cutoff $\Lambda_q b\ll1$, we obtain for $1/2<\alpha<3/2$
\begin{equation}
\langle\langle N^2\rangle\rangle\approx\frac{D}{\pi^2}\int_0^{\Lambda_q}dq\,q^{2\alpha-3}\sin^2(qL/2)[c-(2\alpha-1)\log q],\qquad c=\log(\Lambda_\omega/D).
\end{equation}
The large-$L$ behavior follows by rescaling $x=qL$ and examining the convergence of the resulting integrals.
For $1/2<\alpha<1$, both $\int_0^\infty dx\,x^{2\alpha-3}\sin^2(x/2)$ and its logarithmically weighted counterpart converge, leaving a leading $L^{2-2\alpha}\log L$ contribution.
At $\alpha=1$, the two integrals instead diverge as $\log L$ and $(\log L)^2$, producing a squared logarithmic growth.
For $1<\alpha<3/2$, the integrals grow as powers of the upper cutoff $X=\Lambda_qL$, with their leading logarithmic terms canceling the explicit $\log L$ dependence and leaving a finite, cutoff-dependent constant.
The remaining finite-size correction scales as $L^{2-2\alpha}\log L$.
For $\alpha>3/2$, using the diffusive $q^2$ kernel, direct integration gives the same area-law saturation, with the logarithmic finite-size terms canceling at leading order.
Direct integration therefore yields
\begin{equation}
S_A\propto\langle\langle N^2\rangle\rangle\approx
\begin{cases}
L^{2-2\alpha}\log L, & \frac12<\alpha<1,\\
(\log L)^2, & \alpha=1,\\
C_\infty+O(L^{2-2\alpha}\log L), & 1<\alpha<\frac32,\\
C_\infty+O(L^{-1}), & \alpha>\frac32.
\end{cases}
\end{equation}
where $C_\infty$ is a constant in the thermodynamic limit.
The finite-size corrections are particularly important near $\alpha=1$.
On the area-law side, $L^{2-2\alpha}\log L$ decays slowly when $2\alpha-2$ is small, potentially obscuring saturation even at $L\sim10^3$.
On the sub-volume-law side, the logarithmic factor produces an effective exponent $d\log\langle\langle N^2\rangle\rangle/d\log L\approx2-2\alpha+1/\log L$, which decreases only slowly with system size.

In conclusion, we find the expected marginal behaviour of EE at $\alpha=1$ and the area-law phase for $\alpha>1$, supporting the existence of a metal-insulator phase transition identified by the tree-level RG flow derived from the superdiffusive RA ladder.
Consistent with the established single-particle Anderson localization in PRBM and its connection to the free-fermion many-body ground state, the transition at $\alpha_c=1$ is expected and has already been supported by numerical results~\cite{PouranvariYang2014}, while the expected $(\log L)^2$ scaling at the transition was fitted by a power law $\sim L^{0.33}$, whose small exponent is likely a finite-size artifact because squared-logarithmic scaling is difficult to capture in finite-size simulations.

\section{Long-time entanglement following a unitary quench from high-energy Slater determinant states}
We now turn to the entanglement dynamics in the long-time stationary regime reached through the unitary evolution of an initial high-energy Slater-determinant state, including the N\'eel state used in numerical simulations. It is reasonable to approximate the long-time disorder-averaged state as a maximally mixed state within the fixed-$N$ sector at infinite temperature $\beta=0$. However, the order of disorder averaging and evaluating nonlinear observables matters when we compute properties such as density-density correlations, and directly substituting the disorder-averaged mixed state into Eq.~\ref{eq:SCr} by setting $n_F=\frac12$ incorrectly yields the trivial result $C(r)=0$ for $r>0$, as $\int dE\,A_{ij}(E)=\delta_{ij}$. Moreover, it remains unclear whether the long-distance correlations can be controlled by the diffusive ladder $\Gamma$ when single-particle eigenstates across the entire spectrum contribute. Nevertheless, the density-density correlation function $C(r)$, and thus the leading even particle-number cumulant $\langle\langle N_A^2\rangle\rangle$, provide useful measures of quantum correlations and entanglement, and the latter provides a lower bound on EE, $S_A\geq 4\ln 2\,\langle\langle N_A^2\rangle\rangle$~\cite{GioevKlich2006}, for any pure Slater-determinant state, so that its divergence implies a divergence of EE.

We assume that in the fixed-particle-number $N$ sector in the long-time stationary regime, all occupation configurations of single-particle eigenstates are equally probable for each disorder realization, which yields the maximally mixed state upon occupation averaging, so that $\overline{n_a}=\frac{N}{L}$, $\overline{n_an_b}=\frac{N(N-1)}{L(L-1)}$, $a,b=1,2,\cdots,L$, $a\ne b$, where the average is over different occupations. To show the occupation dependence of $C(r)=\frac{1}{L}\sum_i \overline{\langle c_i^\dag c_{i+r}\rangle \langle c_i c_{i+r}^\dag \rangle}$, we transform to an orthonormal single-particle eigenbasis, $c_i=\sum_{a=1}^L\psi_a(i)d_a$, $c_i^\dag=\sum_{a=1}^L\psi_a^*(i)d_a^\dag$, where $\sum_{i=1}^{L}\psi_a^*(i)\psi_b(i)=\delta_{ab}$, $\sum_{a=1}^L\psi_a(i)\psi_a^{*}(j)=\delta_{ij}$, $\langle d_a^\dag d_b\rangle=\delta_{ab}n_a$, $\langle d_a d_b^\dag\rangle=\delta_{ab}(1-n_a)$, so that
\begin{equation}\label{eq:Br}
C(r)=\frac{1}{L}\sum_{i=1}^L \overline{\sum_{a,b=1}^{L} n_a(1-n_b)\psi_a^*(i)\psi_a(i+r)\psi_b(i)\psi_b^*(i+r)}\overset{L\to\infty}{=}-\frac14 \frac{1}{L}\sum_{i=1}^L\overline{\sum_{a=1}^{L}|\psi_a(i)|^2|\psi_a(i+r)|^2},\quad r>0.
\end{equation}
where we have used the orthonormality and completeness of the eigenfunctions and uniform occupation averaging, and we define $B(r)=\frac{1}{L}\sum_{i=1}^L\overline{\sum_{a=1}^{L}|\psi_a(i)|^2|\psi_a(i+r)|^2}$ as the site- and disorder-averaged long-time single-particle propagation probability. There also exact the exact relation $\langle\langle N^2\rangle\rangle=-\overline{\sum_{i\in A}^{j\in B}\langle c_i^\dag c_j\rangle\langle c_i c_j^\dag\rangle}=-2\sum_{r=1}^{l-1}rC(r)-lC(l)\approx \frac{1}{2}\sum_{r<l}rB(r)$ under periodic boundary condition and the boundary term is subleading in thermodynamic limit.

Compared to the ground state, high-energy states have more available excitation channels and potentially stronger quantum correlations, so for the same hopping exponent $\alpha$, with a fixed single-particle spectrum, they tend toward a more extended entanglement phase. As suggested by numerical results showing that the entanglement transition point shifts to $\alpha>\frac{3}{2}$, we further assume that the transition from area-law entanglement to the extended subvolume-law phase occurs at $\alpha>\frac32$, where the leading long-wavelength dynamics is diffusive, described by $Dq^2-i\omega$. This is somewhat surprising because for $\alpha>\frac32$, the RA ladder does not depend on $\alpha$ at leading order apart from the diffusion coefficient, and subleading effects cannot determine the transition, suggesting that the diffusive pole may no longer provide a small parameter controlling the entanglement transition and that contributions from extended spatial scales become important. This case is similar to the $b\ll1$ regime of single-particle Anderson localization~\cite{MirlinEtAl1996}, where contributions from multiple orbitals are not strongly mixed, because a large $\alpha$ substantially weakens long-range hopping and the associated quantum corrections, so that the small-response-frequency sector described by the diffusive pole cannot alone control the entanglement transition, while the transition may instead be controlled by rare resonances distributed across all spatial scales.

When $\alpha>1$, the single-particle eigenstates are spatially localized, and we consider two localized single-particle reference orbitals $\phi_a$ and $\phi_b$ centered at $x_a$ and $x_b$, with $r=|x_a-x_b|\gg\xi$, where $\xi$ is the localization length and the reference orbitals are defined within the same disorder realization. We define the direct hopping matrix element $v_{ab}=\sum_{i\in a,j\in b}\phi_a^*(i)h_{ij}\phi_b(j)$, whose variance is
\begin{equation}
\overline{|v_{ab}|^2}=\sum_{i\in a,j\in b}|\phi_a(i)|^2\overline{|h_{ij}|^2}|\phi_b(j)|^2\approx\left(\frac{b}{r}\right)^{2\alpha},
\end{equation}
where we have used $\overline{|h_{ij}|^2}\approx(\frac{b}{r})^{2\alpha}$ and the independence and zero mean of the hopping matrix elements, which are also independent of the reference orbitals. The resonance energy window is determined by the root-mean-square hopping matrix element $v(r)\equiv\sqrt{\overline{|v_{ab}|^2}}\approx(\frac{b}{r})^\alpha$. We define $\eta\equiv\nu(E)\int_{\kappa\xi}^{\infty}dr\,v(r)$, where $\nu(E)$ is the density of states and $\kappa\gg1$ is a large constant, as the expected number of direct resonant partners of a localized orbital beyond its localization length, to characterize the sparsity of resonances. We compute $\nu(E)$ directly from the semicircle distribution, as in Eq.~\ref{eq:BS}, obtaining $\nu(E)=\frac{\sqrt{4W_0-E^2}}{2\pi W_0}\sim b^{-\frac12}$, since $W_0\sim b$ and $E$ lies in the bulk of the spectrum. Given $\xi\sim b^2$ for $\alpha>\frac32$, and $\xi\sim b^{(2\alpha-1)/(2\alpha-2)}$ for $\alpha\in(1,\frac32)$, we have
\begin{equation}
\eta\sim b^{\alpha-\frac12}\xi^{1-\alpha}\sim
\begin{cases}
b^{\frac32-\alpha}, & \alpha>\frac32,\\
O(1), & 1<\alpha<\frac32.
\end{cases}
\end{equation}
so that $\eta\to0$ for $\alpha>3/2$ in the large-$b$ limit, supporting the applicability of the rare-resonance approximation, whereas for $1<\alpha<3/2$, $\eta$ is not a small parameter and the approximation is not controlled.

With rare resonances becoming sparse in the regime $\alpha>\frac32$, we consider the resonance between two localized orbitals $\phi_a$ and $\phi_b$ and construct the two-level Hamiltonian $H_{ab}=\begin{pmatrix}E_a & v_{ab} \\ v_{ab}^* & E_b \end{pmatrix}$, with the energy difference $\Delta=E_a-E_b$. Diagonalizing the Hamiltonian and substituting its eigenstates into the expression for $B(r)$ in Eq.~\ref{eq:Br}, we obtain $B_{ij}^{ab}\approx|\phi_a(i)|^2|\phi_b(j)|^2\frac{2|v_{ab}|^2}{\Delta^2+4|v_{ab}|^2}$. Introducing the probability density $p(\Delta)$ of the energy difference and assuming that it is smooth around $\Delta=0$, we integrate over the energy difference
\begin{equation}\label{eq:SCij}
C_{ij}^{ab}\approx \frac{1}{4}|\phi_a(i)|^2|\phi_b(j)|^2
\int d\Delta\,p(\Delta)
\frac{2|v_{ab}|^2}{\Delta^2+4|v_{ab}|^2}
\approx \frac{\pi}{4}|\phi_a(i)|^2|\phi_b(j)|^2
|v_{ab}|p(0).
\end{equation}
where we have approximated $p(\Delta)\approx p(0)$ within the resonance window $|\Delta|\lesssim|v_{ab}|$. Since $v(r)\approx(\frac{b}{r})^\alpha$, $C(r)\sim r^{-\alpha}$, so that the entanglement entropy $S_A\sim\log L$ at $\alpha=2$, which indicates marginal entanglement scaling and supports the entanglement transition.

Then we compute the mutual information $I_2$ and particle-number correlation $G_{AB}$ from the two-level Hamiltonian, where, within the independent-resonance approximation, only pairs connecting regions $A$ and $B$ contribute at leading order. In the eigenbasis, the probability weight of an eigenstate in the state $|\phi_a\rangle$ is $p_a=\frac12(1+\frac{\Delta}{\sqrt{\Delta^2+4|v_{ab}|^2}})$, so for a singly occupied resonance pair, the negative particle-number covariance between the two regions is $G_{AB}^s=p_a(1-p_a)=\frac{|v_{ab}|^2}{\Delta^2+4|v_{ab}|^2}$. For the mutual information, since the isolated pair satisfies $S_{A\cup B}^{ab}=0$, $I_2^s=2S_A=-2p_a\log p_a-2(1-p_a)\log(1-p_a)$, direct integration over the energy difference $\Delta$ gives
\begin{equation}
\frac{I_2}{G_{AB}}=\frac{\int_0^\infty d\Delta\,I_2^s(\Delta)}{\int_0^\infty d\Delta\,G_{AB}^s(\Delta)}=\frac{2\pi|v_{ab}|}{\pi|v_{ab}|/4}=8.
\end{equation}
Here, we use the same approximation as in Eq.~\ref{eq:SCij}, assuming that the probability density is smooth within the narrow resonance window and can be taken outside the integral. We also factor out the common occupation weight. Thus, the coefficient $\frac{2\pi^2}{3}$ predicted by the leading cumulant approximation, $I_2\approx\frac{2\pi^2}{3}G_{AB}$, is replaced by 8, although the entanglement transition point remains governed by the divergence of the spatially weighted sum of the density-density correlator, $\sum_{r=1}^{L}rC(r)\sim\sum_{r=1}^{L}r^{1-\alpha}$, which becomes logarithmic and thus marginal at $\alpha=2$.

In conclusion, although the leading momentum dependence of the retarded-advanced ladder is independent of the hopping exponent $\alpha$ for $\alpha>3/2$, the entanglement in the long-time stationary regime following evolution from a high-energy initial Slater-determinant state is controlled by the exponent $\alpha$ and exhibits an entanglement transition at $\alpha_c=2$. We use a resonance analysis which remains valid in the $\alpha>\frac32$ regime because resonances become sparse due to the strongly decaying long-range hopping, even in the large-bandwidth $b\gg1$ regime. We believe our analysis can be extended across the entire bandwidth range, from $b\ll1$ to $b\gg1$.

\section{Monitored entanglement dynamics by replicated Keldysh formalism}
We begin from the stochastic Schr\"odinger equation of Eq.~\ref{eq:PRBM_QSD} for homodyne measurement, and construct the evolution of the density operator by coupling the bra and the ket, $d\rho_t=|d\psi_t\rangle\langle\psi_t|+|\psi_t\rangle\langle d\psi_t|+|d\psi_t\rangle\langle d\psi_t|$. Generally, the disorder in $\hat{H}_0$ correlates all time points, so the disorder-averaged dynamics is non-Markovian. However, since the hopping disorder is Gaussian, Wick's theorem allows the disorder-averaged Dyson series to be expressed in terms of two-point correlations,
$\overline{\mathcal T\exp\left[-i\sum_{ij}h_{ij}\int_0^t d\tau\,\mathcal L^d_{ij}(\tau)\right]}=\mathcal T\exp\left[-\frac12\sum_{ijkl}\overline{h_{ij}h_{kl}}\int_0^t d\tau\int_0^t d\tau'\,\mathcal L^d_{ij}(\tau)\mathcal L^d_{kl}(\tau')\right]$, where $\mathcal{L}^d_{ij}\rho=[c_i^\dag c_j,\rho]$. The disorder- and monitoring-averaged master equation can then be written as
\begin{equation}
\begin{aligned}
d\overline{\rho_t}={}&-\frac{J^2}{2}dt\sum_{ij}a(|i-j|)\int_0^t d\tau\Big([c_i^\dag(t)c_j(t),[c_i^\dag(\tau)c_j(\tau),\overline{\rho_t}]]+[c_i^\dag(t)c_j(t),[c_j^\dag(\tau)c_i(\tau),\overline{\rho_t}]]\Big)\\
&-\gamma dt\sum_i\int_0^t d\tau\,\delta(t-\tau)[\hat n_i(t),[\hat n_i(\tau),\overline{\rho_t}]].
\end{aligned}
\label{eq:PRBM_master}
\end{equation}
where the symmetrized disorder terms account for both diagonal and off-diagonal hopping matrix elements, and we adopt the endpoint convention $\int_0^t d\tau\,\delta(t-\tau)=1/2$. The entanglement entropy for our Gaussian system can be expanded in a series of particle-number operator cumulants~\cite{KlichLevitov2009}, which are nonlinear functionals of the density matrix, and are therefore represented in the replica formalism by quantities of the form $\Tr(\cdots\rho^{\otimes N})$. Here $\rho$ denotes the unnormalized trajectory density matrix used in the replica representation, so the Born-weighted average contains the factor $(\Tr\rho)^{1-N}$. Following Ref.~\cite{Poboiko2023}, this factor is represented by introducing $R-N$ additional replicas, indexed by $N+1,N+2,\cdots,R$, and finally taking the $R\to1$ limit.
We note that the density matrix in Eq.~(\ref{eq:PRBM_QSD}) is normalized, with the measurement record coupled to the local density deviation $\hat n_i-\langle\hat n_i\rangle_t$. For finite $dt$, the stochastic evolution of the $L\times N$ Slater matrix $U$ requires QR normalization, $U=QR$, where the norm encoded in the triangular factor can be interpreted as the finite-step counterpart of the Born likelihood, providing a numerical connection to the replica limit $R\to1$.

We proceed in the $R$-replicated Hilbert space, $\mathcal{H}^{\otimes R}$, where the monitoring noise $dW_i^t$ and the static Hamiltonian $H_0$ are shared by all $R$ copies. We define the replicated state $|\psi_t\rangle^{\otimes R}=\bigotimes_{a=1}^R|\psi_t\rangle$, and for any operator $\hat A$, its action on replica $a$ is defined by $\hat A^{(a)}=\mathbb I^{\otimes(a-1)}\otimes\hat A\otimes\mathbb I^{\otimes(R-a)}$. Applying the It\^o product rule to the replicated state gives
\begin{equation}
d|\psi_t\rangle^{\otimes R}=\Big[-i\,dt\sum_{a=1}^R H_0^{(a)}+\sum_{i,a}(n_i^{(a)}-\langle n_i\rangle_t)dW_i^t-\frac{\gamma dt}{2}\sum_{i,a}(n_i^{(a)}-\langle n_i\rangle_t)^2+\gamma dt\sum_i\sum_{a<b}(n_i^{(a)}-\langle n_i\rangle_t)(n_i^{(b)}-\langle n_i\rangle_t)\Big]|\psi_t\rangle^{\otimes R}.
\end{equation}
The last term arises from the cross products of the Wiener increments shared by different replicas. The disorder contribution to the replicated evolution is
\begin{equation}
\left.\partial_t\rho^R\right|_{\mathrm{dis}}=-\frac{J^2}{2}\sum_{ij}a(|i-j|)\int_0^t d\tau\sum_{a,b}\Big([c_i^{\dag,(a)}(t)c_j^{(a)}(t),[c_i^{\dag,(b)}(\tau)c_j^{(b)}(\tau),\rho^R(t)]]+[c_i^{\dag,(a)}(t)c_j^{(a)}(t),[c_j^{\dag,(b)}(\tau)c_i^{(b)}(\tau),\rho^R(t)]]\Big)
\label{eq:rho_dis}
\end{equation}
where $\rho^R$ denotes the averaged $R$-replicated density matrix $\rho^R=\overline{\rho^{\otimes R}}$, with the average taken over both the monitoring noise $dW_i^t$ and the static disorder in $H_0$.
The monitoring contribution follows directly from the It\^o product rule, where the single-replica terms give the dephasing dissipator and the shared Wiener increments generate cross-replica couplings,
\begin{equation}
\left.\partial_t\rho^R\right|_{\mathrm{meas}}=-\frac{\gamma}{2}\sum_{i,a}[n_i^{(a)},[n_i^{(a)},\rho^R]]+\gamma\sum_i\sum_{a<b}\left\{n_i^{(a)}-\langle n_i\rangle_t,\left\{n_i^{(b)}-\langle n_i\rangle_t,\rho^R\right\}\right\}
\label{eq:rho_mea}
\end{equation}
where the difference between the double-commutator and double-anticommutator structures in the intra- and inter-replica couplings arises from the Itô drift term $-\frac{\gamma}{2}\sum_{i=1}^{L}\bigl(n_i-\langle n_i\rangle_t\bigr)^2dt$ in Eq.~\ref{eq:PRBM_QSD}.
The terms proportional to $\langle n_i\rangle_t$ are retained because they generate a tadpole contribution in the Keldysh action, while the quadratic $\langle n_i\rangle_t^2$ term does not couple to the fields and can be omitted from that action. The linear tadpole is kept in the subsequent saddle-point and fluctuation analysis. Combining the disorder and monitoring contributions gives $\partial_t\rho^R(t)=\left.\partial_t\rho^R(t)\right|_{\mathrm{dis}}+\left.\partial_t\rho^R(t)\right|_{\mathrm{meas}}$.

Next, we map the replicated master equation to the Keldysh action using the symmetrized (Weyl) ordering, $\hat c_i^\dagger\hat c_j\to\bar\psi_i\psi_j+\frac12\delta_{ij}$, which fixes the equal-time prescription on the Keldysh contour. The Weyl-ordering constants cancel in the disorder commutators, while the monitoring contribution generates both a quartic interaction and a linear term proportional to $(R-1)(1-2\langle n_i\rangle_t)$. We combine the fields in $R$-Replica $\times$ Keldysh space, $\Psi=(\vec\psi^+,\vec\psi^-)$, $\bar\Psi=(\vec{\bar\psi}^+,\vec{\bar\psi}^-)^T$, and obtain
\begin{equation}
\begin{aligned}
iS^{\mathrm{int}}[\bar\Psi,\Psi]={}&\frac{\gamma}{2}\sum_i\int dt\left[\left(\sum_a\bar\Psi_i^a\Psi_i^a\right)^2+2(R-1)(1-2\langle n_i\rangle_t)\sum_a\bar\Psi_i^a\Psi_i^a\right]\\
&-\frac{J^2}{4}\sum_{ij}a(|i-j|)\int dt\,dt'\sum_{a,b}(\bar\Psi_i^a\sigma^z\Psi_j^a)(t)\left[(\bar\Psi_i^b\sigma^z\Psi_j^b)(t')+(\bar\Psi_j^b\sigma^z\Psi_i^b)(t')\right]
\end{aligned}
\label{eq:Psiaction}
\end{equation}
where $\sigma^z$ is the Pauli matrix acting in the $2\times2$ Keldysh space, the factor $i$ in $iS^{\mathrm{int}}$ follows from the real-time Keldysh weight $e^{iS}$, and the superscript $\mathrm{int}$ indicates the exclusion of the free term $iS^{0}[\bar\Psi,\Psi]=i\sum_{i,a}\int dt \bar\Psi_i^a(t)(i\sigma^z\partial_t-0^+\sigma^y)\Psi_i^a(t)$, whose regulator $-0^+\sigma^y$ implements the Keldysh distribution $F(\omega)=1-2n=0$ in the half-filled infinite-temperature steady state. We introduce the doubled basis to combine the two disorder contractions, $\Phi^\dagger=(\bar\Psi,-\Psi)$, $\Phi=(\Psi,\bar\Psi)$, for which Grassmann anticommutation gives $\Phi_i^\dagger\sigma^z\Phi_j=\bar\Psi_i\sigma^z\Psi_j+\bar\Psi_j\sigma^z\Psi_i=\Phi_j^\dagger\sigma^z\Phi_i$. Using $a(|i-j|)=a(|j-i|)$, we obtain
$\sum_{ij}a(|i-j|)(\bar\Psi_i\sigma^z\Psi_j)(t)\left[(\bar\Psi_i\sigma^z\Psi_j)(t')+(\bar\Psi_j\sigma^z\Psi_i)(t')\right]=\frac12\sum_{ij}a(|i-j|)(\Phi_i^\dagger\sigma^z\Phi_j)(t)(\Phi_i^\dagger\sigma^z\Phi_j)(t')$.
The doubled bilinears can then be expressed as site-local matrix products, yielding
\begin{equation}
iS^{\mathrm{int}}[\Phi^\dagger,\Phi]=\Tr\Big[-\frac{\gamma}{8}\delta(t-t')\sum_i(\Phi_i^\dagger\Phi_i)^2+\frac{J^2}{8}\sum_{ij}a(|i-j|)(\Phi_i^\dagger\sigma^z\Phi_i)(\Phi_j^\dagger\sigma^z\Phi_j)+\frac{\gamma(R-1)}{2}\delta(t-t')\sum_i(1-2\langle n_i\rangle_t)(\Phi_i^\dagger\Phi_i)\Big],
\label{eq:SPhiaction}
\end{equation}
where $\Tr$ denotes the matrix trace over the doubled $4R$-dimensional replica--Keldysh space and the convolution over the two time indices $t,t'$. The product of fields is the matrix product (the convolution). Field-independent terms are discarded, while the linear $\langle n_i\rangle_t$ contribution is retained. The free fermionic kernel in the doubled basis is $K_0={\small\begin{pmatrix}i\sigma_K^z\partial_t-0^+\sigma_K^y+i0^+F(\mathbf1_K-\sigma_K^x)&0\\0&-i\sigma_K^z\partial_t+0^+\sigma_K^y+i0^+F(\mathbf1_K-\sigma_K^x)\end{pmatrix}}$ , with $iS_0[\Phi^\dagger,\Phi]=\frac{i}{2}\Tr[\Phi^\dagger K_0\Phi]$, where $F=1-2n$ encodes the prescribed, energy-independent occupation $n$.

Next, we introduce the bilocal auxiliary field $\mathcal G=i\Phi\Phi^\dagger$ through the delta function $1=\int D\mathcal G\delta(\mathcal G-i\Phi\Phi^\dagger)$, and introduce the auxiliary field $Q$ through the corresponding exponential representation,
\begin{equation}\label{eq:HS}
1=\int D\mathcal G DQ\exp\Big(i\Tr\sum_{ij}\left(Q_{ij}(\mathcal G_{ji}-i\Phi_j\Phi_i^\dagger)\right)\Big)=\int D\mathcal G DQ\exp\Big(i\Tr\sum_{ij}\left(Q_{ij}\mathcal G_{ji}+i\Phi^\dagger_i Q_{ij}\Phi_j\right)\Big).
\end{equation}
where we have used the Grassmann anticommutation relation $Q_{ij}\Phi_j\Phi_i^\dagger=-\Phi_i^\dagger Q_{ij}\Phi_j$, with the bosonic collective field $Q$ commuting with the Grassmann fields. Either sign of the $\pm i$ prefactor may be used, and the choice only changes the sign convention for the saddle in the subsequent analysis, $Q_{\mathrm{saddle}}=\pm\sigma_K^z$. We adopt the positive-$i$ convention.

Physically, the collective slow modes carry a small external momentum $q$ and frequency $\Omega$. Consider the four-fermion vertex $\Tr[(\Phi\Phi^\dagger)^2]$, with the fermionic fields ordered as $\Phi_a(k_1)\Phi_b^\dagger(k_2)\Phi_b(k_3)\Phi_a^\dagger(k_4)$ and $k_1-k_2+k_3-k_4=0$. We denote the direct and exchange slow-momentum projections of $\mathcal G$ by $\mathcal G_d$ and $\mathcal G_x$, respectively. The direct pairing with momentum transfer $q_d=k_1-k_2=k_4-k_3$ yields the matrix channel $-\Tr(\mathcal G_d^2)$, whereas the exchange pairing with $q_x=k_1-k_4=k_2-k_3$ yields the trace channel $[\Tr(\mathcal G_x)]^2$. The frequency transfers are defined analogously. In both pairings, the fermionic center momenta are unrestricted and integrated over the full spectrum. The two slow sectors therefore occupy distinct leading regions of the four-leg momentum space. Their overlap, requiring both $q_d$ and $q_x$ to be small, is neglected as subleading in phase space. Before projection, the Grassmann identity $\Tr[(\Phi\Phi^\dagger)^2]=-\Tr(\mathcal G^2)=[\Tr(\mathcal G)]^2$ gives equivalent representations of the full vertex. After projection, however, the Grassmann rearrangement exchanges the constrained momentum transfer and therefore does not commute with the slow-mode projection. Either representation alone retains only one slow sector. We therefore use $\left.\Tr[(\Phi\Phi^\dagger)^2]\right|_{\mathrm{slow}}\simeq-\Tr(\mathcal G_d^2)+[\Tr(\mathcal G_x)]^2$. After integrating over the center momenta, both routings describe the same low-momentum particle-hole fluctuations, so we identify $\mathcal G_d=\mathcal G_x=\mathcal G_{\mathrm{slow}}$ at leading order. The two trace structures, $\Tr\mathcal G$ and $\Tr(\mathcal G\sigma_K^z)$, involve only the replica-singlet components of the Keldysh-diagonal fields and introduce no additional replicon channels. Hereafter, $\mathcal G$ and $Q$ denote their slow-mode projections.

We then derive the slow-mode-projected collective-field interaction action $iS^{\mathrm{int}}[\mathcal G,Q]$ from the doubled fermionic action $iS^{\mathrm{int}}[\Phi^\dagger,\Phi]$. For simplicity, we define $\mathcal G^z=\mathcal G\sigma_K^z$, $Q^z=Q\sigma_K^z$, $g_{ij}(t)=\operatorname{tr}\mathcal G_{ij}(t,t)$, and $g_{ij}^z(t)=\operatorname{tr}\mathcal G_{ij}^z(t,t)$, where $\operatorname{tr}$ acts only on the internal $4R$-dimensional space without temporal integration. We then obtain
\begin{equation}\label{eq:SGQ}
\begin{aligned}
iS^{\mathrm{int}}[\mathcal G,Q]&={}i\Tr\sum_{ij}\mathcal G_{ij}Q_{ji}+\frac{\gamma}{8}\Tr\left[\delta(t-t')\sum_i\mathcal G_{ii}^2\right]-\frac{\gamma}{8}\sum_i\int dt\,dt'\,\delta(t-t')g_{ii}(t)g_{ii}(t')\\
&-\frac{J^2}{8}\Tr\sum_{ij}a(|i-j|)\mathcal G_{ii}^z\mathcal G_{jj}^z+\frac{J^2}{8}\sum_{ij}a(|i-j|)\int dtdt'g_{ij}^z(t)g_{ji}^z(t')+\frac{i\gamma(R-1)}{2}\Tr\Big[\delta(t-t')\sum_i(1-2\langle n_i\rangle_t)\mathcal G_{ii}\Big].
\end{aligned}
\end{equation}
Before performing the Gaussian integration over $\mathcal G$, we separate the quadratic auxiliary-field terms into their equal-time and non-equal-time sectors in the bilocal time-kernel space, $\mathcal G=\mathcal G^{t=t'}+\mathcal G^{t\neq t'}$ and $Q=Q^{t=t'}+Q^{t\neq t'}$. The two sectors have disjoint support and are therefore orthogonal under the full kernel trace. For example, the monitoring quadratic term $\sum_i\int dt\,dt'\,\delta(t-t')\operatorname{tr}[\mathcal G_{ii}(t,t')\mathcal G_{ii}(t',t)]$ is supported only at equal times and reduces to $\sum_i\int dt\,\operatorname{tr}[\mathcal G_{ii}(t,t)^2]$. Consequently, the disorder-induced matrix quadratic term contributes to both the equal-time and non-equal-time sectors, whereas the monitoring terms, the disorder-induced trace-square term, and the tadpole involve only the equal-time components of $\mathcal G$, although the disorder-induced trace-square term couples equal-time components at two independent time points.
We then decompose the interaction action in Eq.~(\ref{eq:SGQ}) into the $t=t'$ and $t\neq t'$ sectors,
\begin{equation}\label{eq:actionS}
\begin{aligned}
iS_{t=t'}^{\mathrm{int}}[\mathcal G^{t=t'},Q^{t=t'}]&={}i\Tr\sum_{ij}\mathcal G_{ij}^{t=t'}Q_{ji}^{t=t'}+\frac{\gamma}{8}\Tr\left[\sum_i(\mathcal G_{ii}^{t=t'})^2\right]-\frac{\gamma}{8}\sum_i\int dt g_{ii}(t)^2
-\frac{1}{8}\Tr\sum_{ij}a(|i-j|)\mathcal G_{ii}^{z,t=t'}\mathcal G_{jj}^{z,t=t'}\\
&+\frac{1}{8}\sum_{ij}a(|i-j|)\int dtdt'g_{ij}^z(t)g_{ji}^z(t')+\frac{i\gamma(R-1)}{2}\Tr\Big[\sum_i(1-2\langle n_i\rangle_t)\mathcal G_{ii}^{t=t'}\Big]\\
iS_{t\neq t'}^{\mathrm{int}}[\mathcal G^{t\neq t'},Q^{t\neq t'}]&={}i\Tr\sum_{ij}\mathcal G_{ij}^{t\neq t'}Q_{ji}^{t\neq t'}-\frac{1}{8}\Tr\sum_{ij}a(|i-j|)\mathcal G_{ii}^{z,t\neq t'}\mathcal G_{jj}^{z,t\neq t'}.
\end{aligned}
\end{equation}
where the tadpole proportional to $g_{ii}(t)$ selects only the equal-time $t=t'$ sector of $\mathcal G$ and set $J=1$.

We first integrate out $\mathcal G^{t\neq t'}$ in the $t\ne t'$ sector. The site-off-diagonal components enter only through the linear coupling to $Q$, so integrating them out sets $Q_{ij}^{t\ne t'}$ to zero for $i\ne j$. The remaining site-diagonal fields form a single Gaussian channel with source $Q_{ii}^{z,t\ne t'}$ and spatial kernel $a(|i-j|)/4$. Performing the Gaussian integral gives
\begin{equation}
iS_{t\ne t'}^{\mathrm{int}}[Q]=-2\operatorname{Tr}\sum_{ij}(a^{-1})_{ij}\,Q_{ii}^{z,t\ne t'}Q_{jj}^{z,t\ne t'},
\end{equation}
where $(a^{-1})_{ij}$ denotes the inverse of the spatial disorder kernel $a(|i-j|)$, and we have omitted a $Q$-independent normalization.

In the $t=t'$ sector we suppress the equal-time superscript for simplicity. For each site pair $i,j$, we extract the $\mathbf 1$ and $\sigma_K^z$ components and split the remaining matrices into Keldysh block-diagonal $(+)$ and block-off-diagonal $(-)$ parts,
\begin{equation}
X_{ij}=X_{+,ij}^{\perp}+X_{-,ij}^{\perp}+\frac{x_{ij}\mathbf 1+x_{ij}^{z}\sigma_K^z}{4R},\quad
\sigma_K^z X_{\pm,ij}^{\perp}\sigma_K^z=\pm X_{\pm,ij}^{\perp},\quad X\in\{\mathcal G,Q\},\quad x\in\{g,q\}
\end{equation}
where $(q_{ij},q_{ij}^{z})=(\operatorname{tr}Q_{ij},\operatorname{tr}Q_{ij}^{z})$.
The normalization follows from $\operatorname{tr}\mathbf 1=\operatorname{tr}[(\sigma_K^z)^2]=4R$, and $\operatorname{tr}(\mathbf 1\,\sigma_K^z)=0$ gives the orthogonality of the two scalar directions. For arbitrary site pairs, the four channels are pairwise orthogonal, $\operatorname{tr}(X_{+,ij}^{\perp}Y_{-,kl}^{\perp})=0,
\operatorname{tr}X_{+,ij}^{\perp}=0, \operatorname{tr}(X_{+,ij}^{\perp}\sigma_K^z)=0,
\operatorname{tr}X_{-,ij}^{\perp}=0, \operatorname{tr}(X_{-,ij}^{\perp}\sigma_K^z)=0$, where $X,Y\in\{\mathcal G,Q\}$.
The two trace conditions make each Keldysh diagonal block of $X_{+,ij}^{\perp}$ traceless and hold automatically for the block-off-diagonal $X_{-,ij}^{\perp}$, so the full block-off-diagonal component satisfies $\mathcal G_-=\mathcal G_-^{\perp}$. Restoring the single time integral in $\operatorname{Tr}$ gives the identities needed for the source and quadratic terms,
\begin{equation}
\begin{aligned}
\operatorname{Tr}(\mathcal G_{ij}Q_{ji})&=\operatorname{Tr}(\mathcal G_{+,ij}^{\perp}Q_{+,ji}^{\perp}+\mathcal G_{-,ij}^{\perp}Q_{-,ji}^{\perp})+\frac{1}{4R}\int dt\,\bigl(g_{ij}q_{ji}+g_{ij}^{z}q_{ji}^{z}\bigr),\\
\operatorname{Tr}(\mathcal G_{ii}\mathcal G_{jj})&=\operatorname{Tr}(\mathcal G_{+,ii}^{\perp}\mathcal G_{+,jj}^{\perp}+\mathcal G_{-,ii}^{\perp}\mathcal G_{-,jj}^{\perp})+\frac{1}{4R}\int dt\,\bigl(g_{ii} g_{jj}+g_{ii}^{z}g_{jj}^{z}\bigr),\\
\operatorname{Tr}(\mathcal G_{ii}^{z}\mathcal G_{jj}^{z})&=\operatorname{Tr}(\mathcal G_{ii}\mathcal G_{jj})-2\operatorname{Tr}(\mathcal G_{-,ii}^{\perp}\mathcal G_{-,jj}^{\perp}),
\end{aligned}
\end{equation}
where scalar fields in a single integral share the time argument $t$. Conjugation by $\sigma_K^z$ reverses the sign of the $(-,\perp)$ component alone. The same quadratic identities hold for $Q$. The integral over $\mathcal G$ therefore factorizes as $\int\mathcal D\mathcal G\,e^{iS^{\mathrm{int}}[\mathcal G,Q]}\propto\left(\int\mathcal D\mathcal G_+^\perp\,e^{iS_+^\perp}\right)\allowbreak\left(\int\mathcal D\mathcal G_-\,e^{iS_-}\right)\allowbreak\left(\int\mathcal Dg\,e^{iS_g}\right)\allowbreak\left(\int\mathcal Dg^z\,e^{iS_{g^z}}\right)$, up to a field-independent Jacobian. The four channel integrals read
\begin{equation}
\begin{aligned}
iS_\pm^\perp
&=\operatorname{Tr}\sum\nolimits_{ij}\left[i\mathcal G_{\pm,ij}^{\perp}Q_{\pm,ji}^{\perp}+\frac{\gamma\delta_{ij}\mp a(|i-j|)}{8}\mathcal G_{\pm,ii}^{\perp}\mathcal G_{\pm,jj}^{\perp}\right],\\
iS_g
&=\sum\nolimits_{ij}\int dt\left[\frac{i}{4R}g_{ij}q_{ji}+\frac{(1-4R)\gamma\delta_{ij}-a(|i-j|)}{32R}g_{ii} g_{jj}\right]+\frac{i\gamma(R-1)}{2}\sum\nolimits_i\int dt\,\bigl(1-2\langle n_i\rangle_t\bigr)g_{ii}(t),\\
iS_{g^z}
&=\sum\nolimits_{ij}\int dt\left[\frac{i}{4R}g_{ij}^{z}q_{ji}^{z}+\frac{\gamma\delta_{ij}-a(|i-j|)}{32R}g_{ii}^{z}g_{jj}^{z}\right]+\frac18\sum\nolimits_{ij}a(|i-j|)\int dt\,dt'\,g_{ij}^{z}(t)g_{ji}^{z}(t').
\end{aligned}
\end{equation}
where the matrix and trace squares from monitoring contribute $\gamma/(32R)$ and $-\gamma/8$ to $g_{ii}^2$, respectively. The tadpole selects $g_{ii}$, and the static disorder trace term couples $g_{ij}^z(t)$ and $g_{ji}^z(t')$ at independent times.

For each independent real eigenmode of the quadratic form in $\mathcal G$, the Gaussian integral converges on the original Hermitian contour for a negative eigenvalue and diverges for a positive eigenvalue.
The divergence requires the analytical continuation of  the corresponding component of $\mathcal G$ from the real axis to the imaginary axis, which can be done as follows.
Taking the  equal-time $(-,\perp)$ channel as an example, we  choose a normalized real spatial eigenmode with $\gamma+a_k>0$, where $a_k=\sum_r a(|r|)e^{-ikr}$, and expand the site-diagonal fields in this mode as  $\mathcal G_-^\perp=\sum_\mu\tilde g_\mu T_\mu$ and $Q_-^\perp=\sum_\mu\tilde q_\mu T_\mu$, where $T_\mu$ form an orthonormal basis of the $-$ channel in the doubled replica$\times$Keldysh space, $\operatorname{tr}(T_\mu T_\nu)=\delta_{\mu\nu}$.
The quadratic kernel of the $\mathcal G$ integral is therefore $K_{\mu\nu}= \frac{\gamma+a_k}{4}\delta_{\mu\nu}$.
For each independent mode, we define the Gaussian integral by analytically continuing the quadratic kernel along the path parametrized by $\phi$,
\begin{equation}
K_{\mu\nu}(\phi)=\frac{\gamma+a_k}{4}e^{i\phi}\delta_{\mu\nu},\qquad \tilde g_\mu=e^{i(\pi-\phi)/2}x_\mu,\qquad \phi:\pi\rightarrow0,
\end{equation}
while keeping the  linear source $\tilde q_\mu$ fixed.
Along this continuation, $\frac12\sum_{\mu,\nu}\tilde g_\mu K_{\mu\nu}(\phi) \tilde g_\nu=-\frac{\gamma+a_k}{8}\sum_\mu x_\mu^2$, so that the convergent real-axis Gaussian at $\phi=\pi$ is continuously continued to the imaginary-axis contour $\tilde g_\mu=ix_\mu$ at $\phi=0$.
Since the integrand is entire in every independent component $\tilde g_\mu$, no finite pole or branch cut is crossed during this continuation.
The auxiliary-field insertion is understood in the finite-$\varepsilon$ regularization introduced in Appendix~B of Ref.~\cite{Poboiko2023}, which fixes the original Fourier representation and normalization of the delta functional.
The analytic continuation described above is a separate convergence prescription for the otherwise divergent $\mathcal G$ Gaussian integral.
It leaves the linear constraint term unchanged and is performed before taking the limit $\varepsilon\to0^+$.

Gaussian integration over $\mathcal G$ gives the four contributions to the equal-time action,
\begin{equation}
\begin{aligned}
iS_{\pm}^\perp[Q]
&=2\operatorname{Tr}\sum_k\frac{Q_{\pm,k}^{\perp}Q_{\pm,-k}^{\perp}}{\gamma\mp a_k},\\
iS_g[Q]
&=\frac{1}{4R}\sum_k\int_{-\infty}^{\infty}\frac{d\omega}{2\pi}\,
\frac{\bigl[q_{ii}(t)+2R\gamma(R-1)(1-2\langle n_i\rangle_t)\bigr]_{k,\omega}
      \bigl[q_{jj}(t)+2R\gamma(R-1)(1-2\langle n_j\rangle_t)\bigr]_{-k,-\omega}}
     {(1-4R)\gamma-a_k},\\
iS_{g^z}[Q]
&=\frac{1}{4R}\sum_k\int_{\omega\ne0}\frac{d\omega}{2\pi}\,
\frac{q_{k,\omega}^zq_{-k,-\omega}^z}{\gamma-a_k}
+\frac{1}{8R^2}\sum_{ij}\frac{q_{ij,0}^zq_{ji,0}^z}{a(|i-j|)},
\end{aligned}
\end{equation}
where $a_k=\sum_r a(|r|)e^{-ikr}$, and $k$ labels the spatial Fourier components of the site-diagonal fields.
The even extension to the full time axis doubles the single-time contraction, giving the prefactor $1/(4R)$ in the scalar frequency integrals.
The matrix trace $\operatorname{Tr}$ includes the original single time integral.
The double-time disorder term depends on $g_{ij}^z(t)$ only through $\int dt\,g_{ij}^z(t)$.
Its conjugate source $q_{ji,0}^z$ is the time-independent component of $q_{ji}^z(t)$, equivalently the coefficient of $2\pi\delta(\omega)$ in $q_{ji,\omega}^z$.
For $i\ne j$, integration over the fields that enter linearly imposes $Q_{\pm,ij}^{\perp}=0$, $q_{ij}=0$, and $q_{ij}^z(t)=q_{ij,0}^z$.
Using the trace identities above and $a_{-k}=a_k$, the four contributions combine to give
\begin{equation}
\begin{aligned}
iS_{t=t'}^{\mathrm{int}}[Q]
={}&2\operatorname{Tr}\sum_k\left[
\frac{\gamma Q_kQ_{-k}+a_kQ_k^zQ_{-k}^z}{\gamma^2-a_k^2}
-\frac{q_{k,0}^z}{4R(\gamma-a_k)}
 \left(2Q_{-k}^z-\frac{q_{-k,0}^z}{4R}\mathbf 1\right)
\right]\\
&+\frac{1}{8R^2}\sum_{ij}\frac{q_{ij,0}^zq_{ji,0}^z}{a(|i-j|)}
+\gamma\sum_k\int_{-\infty}^{\infty}\frac{d\omega}{2\pi}\,
\frac{q_{k,\omega}q_{-k,-\omega}}
     {(\gamma-a_k)\bigl[(1-4R)\gamma-a_k\bigr]}\\
&+\gamma(R-1)\sum_k\int_{-\infty}^{\infty}\frac{d\omega}{2\pi}\,
\frac{[1-2\langle n_i\rangle_t]_{k,\omega}
      \bigl(q_{-k,-\omega}+R\gamma(R-1)[1-2\langle n_j\rangle_t]_{-k,-\omega}\bigr)}
     {(1-4R)\gamma-a_k},
\end{aligned}\label{eq:Qequal}
\end{equation}
where $q_{k,0}^z=L^{-1/2}\sum_i e^{-iki}q_{ii,0}^z$.
Under the $k$ sum, the subtraction inside $\operatorname{Tr}$ leaves the scalar product $(q_k^z(t)-q_{k,0}^z)(q_{-k}^z(t)-q_{-k,0}^z)$ with single-time weight $1/[2R(\gamma-a_k)]$.
The sum over site pairs thus supplies the entire zero-frequency weight in the $g^z$ channel, including $i=j$.

We now determine the physical saddle of the equal-time action. We take $R=1$ and seek a stationary, spatially homogeneous, site-diagonal saddle $Q_{ij}(t)=\delta_{ij}Q_s$ in the number-conserving sector. The terms proportional to $R-1$ therefore drop out. We take the saddle to be local in time, so $Q_s^{t\ne t'}=0$ and the entire interaction term $S^{\mathrm{int}}_{t\ne t'}[Q_s]$ vanishes. The fermionic trace logarithm is evaluated at the equal-time saddle. The equal-time causality condition reads $q_{ii}^z(t)=0$ in the contour basis. In the combined result of the four Gaussian integrals, this allows us to omit the diagonal zero-frequency subtraction and the $i=j$ contribution to the static term. We retain the ordinary trace $q_{ii}(t)$ in the combined action.

To separate the retarded and advanced diagonal components, we perform the Keldysh rotation using the convention $\Psi_{\pm}=\tilde U\Psi_K$ and $\bar\Psi_{\pm}=\bar\Psi_K\tilde U\sigma_K^z$, with $\tilde U=\frac{1}{\sqrt{2}}\begin{pmatrix}1&1\\1&-1\end{pmatrix}$. Accordingly, a bilinear kernel transforms as $A_K=\tilde U\sigma_K^zA_{\pm}\tilde U$, so that $\mathbf1_K\to\sigma_K^x$ and $\sigma_K^z\to\mathbf1_K$. The scalar components are $q_{ij}=\operatorname{tr}(Q_{K,ij}\sigma_K^x)$ and $q_{ij}^z=\operatorname{tr}Q_{K,ij}$. We henceforth omit the Keldysh subscript on $Q$ and $K_0$. The dressed Green function is $G[Q]=(K_0+2iQ)^{-1}$, with $G_s=G[Q_s]$ and $Q_s$ taking an upper-triangular form in Keldysh space,
$G_s=\begin{pmatrix}G_s^R&G_s^K\\0&G_s^A\end{pmatrix},
Q_s=\begin{pmatrix}Q_s^R&Q_s^K\\0&Q_s^A\end{pmatrix}$.

At $R=1$, applying the $2\times2$ trace identity in each of the two transpose-related normal blocks gives $\operatorname{Tr}(\sigma_K^xQ_k\sigma_K^xQ_{-k})=-\operatorname{Tr}(Q_kQ_{-k})+\frac12\int dt\,q_k(t)q_{-k}(t)$, where we have used $q_{ii}^z(t)=0$. The factor $1/2$ arises because $q_k$ already includes both blocks. Using the full-line frequency convention of the Gaussian integrals, the resulting action before rescaling $Q$ is
\begin{equation}
iS[Q]=\frac12\operatorname{Tr}\sum_i\bigl[\ln(K_0+2iQ)\bigr]_{ii}-2\operatorname{Tr}\sum_k\frac{Q_kQ_{-k}}{\gamma+a_k}+\frac{\gamma}{2}\sum_k\int_{-\infty}^{\infty}\frac{d\omega}{2\pi}\frac{q_{k,\omega}q_{-k,-\omega}}{(\gamma+a_k)(3\gamma+a_k)}+\frac18\sum_{i\ne j}\frac{q_{ij,0}^zq_{ji,0}^z}{a(|i-j|)},
\end{equation}
where $Q_k$ is the spatial Fourier component of $Q_{ii}$ and $q_{ij,0}^z$ is the zero-frequency amplitude of $q_{ij}^z(t)$.

We now use the half-filling condition $n=1/2$ of the homogeneous infinite-temperature steady state. Its energy-independent distribution $F=1-2n=0$ gives $G_s^K=0$. Since the Keldysh component of $K_0$ also vanishes, the relation $G_s^{-1}=K_0+2iQ_s$ gives $Q_s^K=0$ and hence $q_{ii}=0$. The remaining static term vanishes because $q_{ij,0}^z=0$ for $i\ne j$ at the site-diagonal saddle. The action for determining this saddle therefore reduces to
\begin{equation}
iS[Q]=\frac12\operatorname{Tr}\sum_i\bigl[\ln(K_0+2iQ)\bigr]_{ii}-2\operatorname{Tr}\sum_k\frac{Q_kQ_{-k}}{\gamma+a_k},
\end{equation}
where the variations of the ordinary-trace and static terms vanish at $q_{ii}=0$ and $q_{ij,0}^z=0$ for $i\ne j$, respectively. Away from half filling, $n\neq \frac12$, the ordinary-trace channel contributes an additional term to the saddle-point action and nonzero Keldysh components must be retained, while the retarded and advanced saddle-point equations are expected to remain decoupled and yield the standard SCBA solution.

Varying the reduced action and imposing the homogeneous saddle-point condition $Q_i=Q$ gives $i\int_{-\infty}^{\infty}\frac{d\varepsilon}{2\pi}\,G_s(\varepsilon)-\frac{4Q_s}{\gamma+a_{k=0}}=0$, where $a_{k=0}=\sum_r a(|r|)$. The retarded and advanced components give two independent equations,
\begin{equation}
\frac{4Q_s^R}{\gamma+a_{k=0}}=i\int_{-\infty}^{\infty}\frac{d\varepsilon}{2\pi}\frac{1}{\varepsilon+i0^++2iQ_s^R}=\frac12,\quad
\frac{4Q_s^A}{\gamma+a_{k=0}}=i\int_{-\infty}^{\infty}\frac{d\varepsilon}{2\pi}\frac{1}{\varepsilon-i0^++2iQ_s^A}=-\frac12,
\end{equation}
where the energy integrals are taken symmetrically about $\varepsilon=0$. The two $2\times2$ blocks of $K_0+2iQ_s$ in the doubled basis are related by $\varepsilon\to-\varepsilon$ and hence yield identical energy integrals. The resulting factor of two is common to both terms of the saddle-point equation and therefore cancels. Thus $Q_s^R+Q_s^A=0$. The saddle of the original auxiliary field is therefore $Q_s=(\gamma+a_{k=0})\sigma_K^z/8$.

Finally, we take $8Q/(\gamma+a_{k=0})$ as the integration variable and denote it again by $Q$. The normalized saddle is $Q_s=\sigma_K^z$, and the rescaled action including the $t\ne t'$ interaction term reads
\begin{equation}
iS[Q]=\frac12\operatorname{Tr}\sum_i\left[\ln\left(K_0+\frac{i(\gamma+a_{k=0})}{4}Q\right)\right]_{ii}-\frac{(\gamma+a_{k=0})^2}{32}\operatorname{Tr}\sum_k\left(\frac{Q_kQ_{-k}}{\gamma+a_k}+\frac{Q_k^{t\ne t'}Q_{-k}^{t\ne t'}}{a_k}\right),
\end{equation}
where the trace-log term includes both temporal sectors and the omitted scalar terms vanish for the soft-mode configurations considered below. The corresponding inverse Green function is $G_s^{-1}(\varepsilon)={\small\begin{pmatrix}\varepsilon+i[0^++(\gamma+a_{k=0})/4]\sigma_K^z&0\\0&-\varepsilon+i[0^++(\gamma+a_{k=0})/4]\sigma_K^z\end{pmatrix}}$.

Since the saddle is supported in the equal-time sector, we have $(Q^{t=t'})^2\approx\mathbf1$ in the leading saddle-point approximation, whereas $(Q^{t\ne t'})^2$ is quadratic in the fluctuations and cannot be reduced to a constant. Using the constraint $Q^2=\mathbf1$ for the full field, the common momentum-independent term is constant, while the difference between the two coefficients in the long-wavelength limit yields $-\frac{\gamma(\gamma+a_{k=0})}{32a_{k=0}}\operatorname{Tr}\sum_k Q_k^{t\ne t'}Q_{-k}^{t\ne t'}$. The trace-log term is invariant under stationary and spatially homogeneous rotations on the saddle manifold in the zero limit. Thus, the non-equal-time fluctuations acquire a finite nonderivative quadratic term for $\gamma>0$. Below the corresponding mass scale, these fluctuations have no infrared singularity and can be integrated out. Their integration may renormalize the bare NLSM couplings, but does not control the infrared RG flow of the remaining equal-time sector. We therefore omit this massive sector from the subsequent RG analysis.

We now extract the homogeneous part of the spatial interaction by writing $(\gamma+a_k)^{-1}=(\gamma+a_{k=0})^{-1}+(a_{k=0}-a_k)/[(\gamma+a_{k=0})(\gamma+a_k)]$. The first term yields a constant due to the constraint $Q^2=\mathbf1$. In the continuum and long-wavelength limits, $b\gg1$, $L\gg b$ and $|k|b\ll1$, we have $a_{k=0}\sim b$, since both the central band and the power-law tail contribute linearly in $b$ for $\alpha>1/2$. For $\alpha\in(1/2,3/2)$, we have $a_{k=0}-a_k=2\sum_{r=1}^{\infty}a(r)(1-\cos kr)\approx C_\alpha b^{2\alpha}|k|^{2\alpha-1}$. Since $(a_{k=0}-a_k)/(\gamma+a_{k=0})=O((|k|b)^{2\alpha-1})$, we approximate $\gamma+a_k\approx\gamma+a_{k=0}$ in the denominator of the second term. The resulting action reads
\begin{equation}\label{eq:Qlong}
iS[Q]\approx\frac12\operatorname{Tr}\sum_i\left[\ln\left(K_0+\frac{i(\gamma+a_{k=0})}{4}Q\right)\right]_{ii}-\frac{C_\alpha b^{2\alpha}}{32}\operatorname{Tr}\sum_k|k|^{2\alpha-1}Q_kQ_{-k},
\end{equation}
where $C_\alpha=2\int_0^\infty dx \frac{1-\cos x}{x^{2\alpha}}$.

At leading order, we characterize the entanglement entropy through the second cumulant of the particle number in a subregion $A$, $C_A^{(2)}=\overline{\langle(\delta N_A)^2\rangle}=\sum_{i,j\in A}C_{ij}$, where $C_{ij}=\overline{\langle\delta n_i\delta n_j\rangle}$ and $N_A=\sum_{i\in A}n_i$. We decompose the theory into the replica-symmetric and replicon sectors. In the replica-symmetric sector associated with the global $U(1)$ symmetry, we expect the density dynamics to be governed by the retarded-advanced ladder $K(q,\Omega)\sim(D|q|^{2\alpha-1}-i\Omega)^{-1}$, with $D\sim b^{2\alpha}$. Its time-domain propagation decays as $K(q,t)\sim e^{-D|q|^{2\alpha-1}t}$ for nonzero $q$. Thus, the replica-symmetric contribution vanishes in the long-time limit and does not contribute to the steady-state connected density correlation.

The steady-state connected density correlation is determined by the replica-traceless density response. Taking two derivatives of the generating functional with respect to a replica-traceless density source gives $C_{\mathrm{repl}}(q;t,t')=\frac{1}{4\gamma}\delta(t-t')-\frac{1}{16\gamma^2}\partial_t\partial_{t'}\Gamma(q;t,t')$, where $\Gamma(q;t,t')$ denotes the Gaussian propagator in the replicon sector. The physical equal-time correlation follows by taking $t=t'=t_f$, $C(q)=C_{\mathrm{repl}}(q;t_f,t_f)$. We therefore expand the action around the saddle point to derive the replicon propagator $\Gamma$.

We now analyze the symmetry of the replicated action in Eq.~\ref{eq:SPhiaction} to characterize the allowed transformations of the fields. The doubled basis $\Phi^\dagger,\Phi$ does not introduce additional independent fermionic degrees of freedom or enlarge the symmetry of the original action in terms of $\bar\Psi,\Psi$. Under the unitary transformation $\Phi\to\mathcal U\Phi$, $\Phi^\dagger\to\Phi^\dagger\mathcal U^\dagger$, the two bilinear channels $\Phi^\dagger\Phi$ and $\Phi^\dagger\sigma^z\Phi$, in the original Keldysh $\pm$ basis, jointly require $\mathcal U^\dagger\mathcal U=\mathbf 1$ and $\mathcal U^\dagger\sigma^z\mathcal U=\sigma^z$, yielding the symmetry group $G=U(R)\times U(R)$. After the Keldysh rotation, the homogeneous saddle $\Lambda=\sigma_K^z$ spontaneously breaks $G$ to the stabilizer subgroup $H=\Delta U(R)\simeq U(R)$, defined by $H=\{g\in G\mid g\Lambda g^{-1}=\Lambda\}$. Physically, the broken symmetry generates a continuous family of degenerate saddle configurations, while transformations in $H$ leave the reference saddle unchanged. The distinct saddle configurations therefore form the coset $G/H\simeq U(R)$, whose Goldstone modes constitute the full saddle manifold. Separating the replica-symmetric $U(1)$ mode leaves the $SU(R)$ replicon sector governing the steady-state entanglement dynamics.

For the replicon sector, we parametrize the field and its linear fluctuation around the fixed homogeneous solution $\Lambda\equiv\sigma_K^z$ as
\begin{equation}
Q=T\Lambda T^{-1},\qquad T=e^{i\Xi\sigma_K^x/2},\qquad i\Xi\in\mathfrak{su}(R),\qquad \delta Q=\frac{i}{2}\Xi[\sigma_K^x,\Lambda]=\Xi\sigma_K^y.
\end{equation}
In the number-conserving sector, the two diagonal $2R\times2R$ kernels acting on the $\Psi$ and $\bar\Psi$ components of the doubled field are related by transposition of both the internal indices and the time arguments. Their trace logarithms coincide, so evaluating the doubled trace cancels the original prefactor $1/2$. Cyclically moving $T$ under the trace fixes the auxiliary-field contribution at $\Lambda$ and generates $T^{-1}[i\partial_t,T]=ia_t$, where $a_t=T^{-1}\partial_tT$. We subtract the field-independent value and take the zero limit of the regulator. We assume that $T$ varies slowly in time, such that the external frequency $\Omega$ carried by the collective field satisfies $|\Omega|\ll(\gamma+a_{k=0})/4$. Using $G_s$, the dressed Green function evaluated at $\Lambda$, we expand to second order in time derivatives while retaining the nonlinear dependence on $T$,
\begin{equation}
\operatorname{Tr}\sum_i\bigl[\ln(\mathbf1+iG_sa_t)\bigr]_{ii}\approx i\operatorname{Tr}\sum_i[G_sa_t]_{ii}+\frac12\operatorname{Tr}\sum_i[G_sa_tG_sa_t]_{ii}.
\end{equation}
We now introduce $U=e^{i\Xi}\in SU(R)$ and use the projectors $P_\pm^x=(\mathbf1_K\pm\sigma_K^x)/2$ to write $T=P_+^xU^{1/2}+P_-^xU^{-1/2}$, so that $a_t=P_+^xU^{-1/2}\partial_tU^{1/2}+P_-^xU^{1/2}\partial_tU^{-1/2}$. Using the previously evaluated energy integral, the linear term becomes $i\operatorname{Tr}\sum_i[G_sa_t]_{ii}=\frac12\operatorname{Tr}\sum_i[\Lambda a_t]_{ii}=0$, since the Keldysh traces of $\Lambda P_\pm^x$ vanish.

For the quadratic term, with the doubled-basis multiplicity accounted for above, we decompose $G_s=P_+G_s^R+P_-G_s^A$, where $P_\pm=(\mathbf1_K\pm\Lambda)/2$. The closed $RR$ and $AA$ contributions vanish because a closed loop of purely retarded or purely advanced Green functions cannot satisfy causal time ordering, leaving only the $RA$ and $AR$ contributions~\cite{kamenev2011}. At leading order in the external frequency, we use $G_s^{R,A}(\varepsilon+\Omega)\approx G_s^{R,A}(\varepsilon)$. The corresponding coefficient is $D_t=\int_{-\infty}^{\infty}\frac{d\varepsilon}{2\pi}G_s^R(\varepsilon)G_s^A(\varepsilon)=\int_{-\infty}^{\infty}
\frac{d\varepsilon}{2\pi}\bigl[\varepsilon^2+(\gamma+a_{k=0})^2/16\bigr]^{-1}=2/(\gamma+a_{k=0})$. 
Using the projector identity $\operatorname{Tr}(P_+a_tP_-a_t+P_-a_tP_+a_t)=-\frac14\operatorname{Tr}[\Lambda,a_t]^2$, we obtain
\begin{equation}
\operatorname{Tr}\sum_i[G_sa_tG_sa_t]_{ii}\approx-\frac{D_t}{4}\operatorname{Tr}\sum_i\bigl([\Lambda,a_t]^2\bigr)_{ii}.
\end{equation}
The commutator can be written as $[\Lambda,a_t]=i\sigma_K^yU^{-1/2}(\partial_tU)U^{-1/2}$. Using $U^\dagger=U^{-1}$, the Keldysh trace and cyclicity give $\operatorname{Tr}\sum_i\bigl([\Lambda,a_t]^2\bigr)_{ii}=2\operatorname{Tr}\sum_i(\partial_tU_i^\dagger)(\partial_tU_i)$. The temporal contribution is therefore $-\frac{D_t}{4}\operatorname{Tr}\sum_i(\partial_tU_i^\dagger)(\partial_tU_i)$.

For the spatial term, we evaluate each of the two diagonal contributions in the original doubled trace using $Q=P_+^x\Lambda P_-^xU+P_-^x\Lambda P_+^xU^\dagger$. Since $P_\pm^x\Lambda=\Lambda P_\mp^x$, the product becomes
\begin{equation}
Q_kQ_{-k}=P_+^xU_k(U^\dagger)_{-k}+P_-^x(U^\dagger)_kU_{-k}.
\end{equation}
The Keldysh trace of each projector is unity, and the two terms contribute equally after summing over $k$ with the even kernel $|k|^{2\alpha-1}$. Together with the equal contributions from the two diagonal kernels in the doubled basis, this gives a total factor of four in the spatial term. Combining the temporal and spatial contributions, we obtain the $SU(R)$ NLSM in the superdiffusive regime $\alpha\in(1/2,\frac32)$,
\begin{equation}
iS[U]\approx-\operatorname{Tr}\left[\frac{1}{2(\gamma+a_{k=0})}\sum_i(\partial_tU_i^\dagger)(\partial_tU_i)+\frac{C_\alpha b^{2\alpha}}{8}\sum_k|k|^{2\alpha-1}U_k(U^\dagger)_{-k}\right].
\end{equation}

The two coefficients can be made equal by rescaling time, $\tilde t=\frac{b^\alpha}{2}\sqrt{C_\alpha(\gamma+a_{k=0})}\,t$. Transforming both the time derivatives and the time integration contained in $\operatorname{Tr}$, and Fourier transforming with respect to $\tilde t$, we obtain
\begin{equation}
iS[U]\approx-\frac{b^\alpha}{4}\sqrt{\frac{C_\alpha}{\gamma+a_{k=0}}}\operatorname{Tr}\sum_k\left(|k|^{2\alpha-1}+\tilde\Omega^2\right)U_{k,\tilde\Omega}(U^\dagger)_{-k,-\tilde\Omega}.
\end{equation}
Here $\tilde\Omega$ is conjugate to $\tilde t$, and its integration is included in $\operatorname{Tr}$.

We perform a tree-level scaling analysis of the NLSM kernel $|k|^{2\alpha-1}+\tilde\Omega^2$. We identify the prefactor of the rescaled action, $g\equiv\frac{b^\alpha}{4}\sqrt{\frac{C_\alpha}{\gamma+a_{k=0}}}$, as the bare dimensionless stiffness. Under the anisotropic rescaling $x\to sx$ and $\tilde t\to s^z\tilde t$, balancing the two terms, $1-2\alpha=-2z$, gives the dynamical exponent $z=\alpha-\frac12$. Since the constrained field $U$ remains dimensionless, the stiffness transforms as $g\to s^{1-z}g$, yielding the RG flow
\begin{equation}
\frac{dg}{d\log l}=\left(\frac32-\alpha\right)g,
\end{equation}
where $l$ denotes the running RG length, $l_0$ is the initial coarse-graining length, and $g(l_0)=g$ is the bare stiffness. For $1<\alpha<\frac32$, the positive RG coefficient drives the stiffness toward larger values, corresponding to a flow toward the weak-coupling regime of the NLSM. At $\alpha=\frac32$, the stiffness becomes marginal at tree level, identifying the upper boundary of the superdiffusive regime. This boundary is reached continuously from within $1<\alpha<\frac32$, as the spatial kernel approaches a quadratic momentum dependence and the dynamical exponent approaches $z=1$. For $\alpha>\frac32$, although the quadratic spatial kernel describes diffusive propagation at short times, the negative RG coefficient drives the stiffness toward smaller values, corresponding to a flow toward the strong-coupling regime of the NLSM, which is associated with area-law entanglement.

Near the marginal boundary, expanding the tree-level RG solution gives $g(L)/g(l_0)=\exp[(\frac32-\alpha)\log(L/l_0)]=1+(\frac32-\alpha)\log(L/l_0)+\cdots$, demonstrating that the finite-size scaling is governed by a logarithmic dependence on system size, in contrast to the conventional power-law scaling $L^{1/\nu}$. Accordingly, when the data are analyzed using standard polynomial fitting with scaling fields $\phi_i=u_i(w)L^{1/\nu}$, where $u_i(w)$ is expanded polynomially in the reduced hopping exponent $w=(\alpha-\alpha_c)/\alpha_c$, the fitted effective exponent $\nu$ increases with system size rather than converging to a finite value.

We further characterize the entanglement scaling through the connected density correlation, which follows from the replicon propagator through two time derivatives. In Fourier space, the two derivatives give a factor of $\tilde\Omega^2$, and the cancellation of the momentum-independent contact contribution leaves the frequency integral $C(q)\sim\int\frac{d\tilde\Omega}{2\pi}\left[\frac{\tilde\Omega^2}{\tilde\Omega^2+|q|^{2\alpha-1}}-1\right]\sim-|q|^{\alpha-\frac12}$, yielding $C(r)\sim r^{-(\alpha+\frac12)}$ for $1<\alpha<\frac32$. This power law continuously approaches $C(r)\sim r^{-2}$ at the marginal boundary, where the corresponding second cumulant becomes logarithmic, giving logarithmic entanglement scaling. Meanwhile, $G_{AB}$ remains scale invariant because the two spatial integrations contribute $L^2$, which precisely compensates the $L^{-2}$ decay of the critical density correlation. The mutual information $I_2$ shares this scale invariance through its leading-order relation to $G_{AB}$. For $\alpha>\frac32$, the critical algebraic correlation no longer persists at long distances, and the area-law entanglement saturates to a constant, while $G_{AB}$ and $I_2$ vanish with increasing system size as the two macroscopically separated regions become asymptotically uncorrelated.

In conclusion, we identify an entanglement transition at $\alpha_c=\frac32$ in the PRBM under continuous homodyne monitoring, separating the subvolume-law and area-law entanglement phases. To explain this transition, we derive an effective superdiffusive nonlinear sigma model incorporating an additional Euclidean time dimension, whose interplay with the long-range spatial kernel renders the transition marginal at tree level. The critical point is continuously approached from the superdiffusive regime $1<\alpha<\frac32$ and coincides with the crossover to quadratic spatial dispersion, which describes short-time diffusive propagation. Our analysis applies in the large-bandwidth and weak-monitoring regime, $b\gg1$ and $\gamma\ll J\sqrt b$.
\end{document}